\documentclass[aip, jcp, notitlepage, floatfix, reprint, citeautoscript, twocolumn]{revtex4-1}

\usepackage{amsmath}
\usepackage{amsfonts}
\usepackage{graphicx}
\usepackage{dcolumn}
\usepackage[table]{xcolor}
\usepackage[version=3]{mhchem}
\usepackage{transparent}
\usepackage{gensymb}
\usepackage{tabularx}
\usepackage{etoolbox}
\usepackage[hidelinks]{hyperref}
\usepackage{bm}
\usepackage{relsize}
\usepackage[normalem]{ulem}

\def\maxfloatwidth{%
  \ifdim\columnwidth>246.0pt
  300.0pt  \else
  \columnwidth
  \fi
}

\newcommand{\tbf}[1]{\textbf{#1}}

\newcommand{\mrm}[1]{\mathrm{#1}}
\newcommand{\mbf}[1]{\mathbf{#1}}

\newcommand{\etal}{\emph{et al.}}
\newcommand{\me}{\mathrm{e}}
\newcommand{\dctfac}{\bigg(\frac{\epsilon-1}{\epsilon}\bigg)}

\newcommand{\bohr}{$a_{0}$}

\begin{document}

\title{Interfacial ion solvation: Obtaining the thermodynamic limit
  from molecular simulations}

\author{Stephen J. Cox} 
\affiliation{Chemical Sciences Division, Lawrence Berkeley National
  Laboratory, Berkeley, CA 94720, United States.}

\author{Phillip L. Geissler}
\affiliation{Chemical Sciences Division, Lawrence Berkeley National
  Laboratory, Berkeley, CA 94720, United States.}
\affiliation{Department of Chemistry, University of California,
  Berkeley, CA 94720, United States.}
\email{geissler@berkeley.edu}

\date{\today}

\begin{abstract}
Inferring properties of macroscopic solutions from molecular
simulations is complicated by the limited size of systems that can be
feasibly examined with a computer. When long-ranged electrostatic
interactions are involved, the resulting finite size effects can be
substantial and may attenuate very slowly with increasing system size,
as shown by previous work on dilute ions in bulk aqueous solution.
Here we examine corrections for such effects, with an emphasis on
solvation near interfaces. Our central assumption follows the
perspective of H\"{u}nenberger and McCammon [J. Chem. Phys. \tbf{110},
  1856 (1999)]: Long-wavelength solvent response underlying finite
size effects should be well described by reduced models like
dielectric continuum theory, whose size dependence can be calculated
straightforwardly. Applied to an ion in a periodic slab of liquid
coexisting with vapor, this approach yields a finite size correction
for solvation free energies that differs in important ways from
results previously derived for bulk solution. For a model polar
solvent, we show that this new correction quantitatively accounts for
the variation of solvation free energy with volume and aspect ratio of
the simulation cell. Correcting periodic slab results for an aqueous
system requires an additional accounting for the solvent's intrinsic
charge asymmetry, which shifts electric potentials in a size-dependent
manner. The accuracy of these finite size corrections establishes a
simple method for \textit{a posteriori} extrapolation to the
thermodynamic limit, and also underscores the realism of dielectric
continuum theory down to the nanometer scale.
\end{abstract}

\maketitle

\section{Introduction}
\label{sec:intro}

Molecular simulations of ions near the boundaries of liquid solutions
\cite{jungwirth2001molecular,jungwirth2002ions,jungwirth2006specific,ben2016interfacial,baer2011toward,tobias2013simulation,netz2012progress,levin2009polarizable,levin2009ions,OttenSaykally2012sjc,archontis2006dissecting,arslanargin2012free,baer2014toward,beck2011local,mccaffrey2017mechanism,caleman2011atomistic,herce2005surface,kumar2013exploring,levin2014ions,ou2013spherical,ou2013temperature,stern2013thermodynamics,tse2015propensity,vaikuntanathan2013adsorption,whitmer2014surface,yagasaki2010effects,baer2012electrochemical,Mucha2005unified,petersen2005enhanced,archontis2005attraction,Noah-VanhouckeGeissler2009sjc}
have challenged and reshaped traditional microscopic perspectives on
solvation thermodynamics \cite{onsager1934surface}. In the case of
aqueous interfaces they have inspired a new generation of experiments
\cite{OttenSaykally2012sjc,piatkowski2014extreme,petersen2004confirmation,PetersenSaykally2006sjc,verreault2012conventional,mishra2012anions,liu2004vibrational,Mucha2005unified,raymond2004probing,tarbuck2006spectroscopic,viswanath2007oriented,shultz2000sum,petersen2005enhanced,mccaffrey2017mechanism,weber2004photoemission}
and theories
\cite{archontis2005attraction,vrbka2004propensity,levin2009polarizable,levin2009ions,levin2014ions,OttenSaykally2012sjc,vaikuntanathan2013adsorption,ben2016interfacial}
to identify basic driving forces and their implications across a broad
range of fields, such as electrochemistry
\cite{schmickler2010interfacial}, aerosols
\cite{finlayson2003tropospheric,knipping2000experiments} and protein
surfaces \cite{baldwin1996hofmeister,jungwirth2008ions}.  The great
strength of such simulations---resolving atomistically detailed
structure on ultrafast time scales---is counterbalanced, however, by
the necessary use of imperfect potential energy models and by the
necessary restriction to small system sizes. This paper addresses the
latter problem, aiming to extract from microscopically finite
simulations the behavior of macroscopically large solutions and their
interfaces.

Efforts to correct for finite size effects in molecular simulation are
nearly as old as the simulation methods themselves. Most notably, the
use of periodic boundary conditions (PBC) mitigates the
unrealistically high surface-to-volume ratios, and degrees of
interfacial curvature of small systems. Even for systems that interact
with relatively short-ranged interactions, however, careful treatment
of interactions between distant particles is often required to obtain
well-converged thermodynamic properties
\cite{allen1987computer,lague2004pressure,ou2005molecular,shirts2007accurate,shen2007comparative,martinez2014effect,fischer2015properties,lundberg2016dispersion,ghoufi2017importance,chen1997fast,ko2002rapidly,shi2006molecular,in2007application,wennberg2013lennard,wennberg2015direct}.
In the case of ion solvation, the very slow decay of Coulomb
interactions with distance demands an even more careful
treatment. Indeed, a simulation cell bearing a net charge cannot be
periodically replicated without incurring a super-extensive energetic
cost. Whether implicitly or explicitly, neutralizing countercharge
must be introduced to ensure a well-defined thermodynamic limit when
an ion is subject to periodic boundary conditions. Here we will
exclusively consider systems that are periodic in three dimensions,
whose unit cells contain a single charged solute and a spatially
uniform neutralizing background charge density, in addition to a
collection of electroneutral solvent molecules. Due to the long range
of electrostatic forces, periodic replicas of these charged
constituents can generate substantial fields and potentials, which in
turn induce solvent response that decays only gradually with
increasing system size. These are the contributions we wish to assess
and correct.

While far from macroscopic, computationally accessible systems are
large enough to capture much of the complicated molecular-scale
response that is missing from the classic linear response theory for
solvent polarization, i.e., dielectric continuum theory (DCT). Our
approach asserts that DCT accurately describes solvent response at all
larger scales. As such, finite size effects in DCT should serve as
useful estimates for those of more complicated molecular models. This
perspective was employed by H\"{u}nenberger and McCammon
\cite{HunenbergerMcCammon1999sjc} to investigate the nature and
magnitude of periodicity-induced artifacts for bulk solutions,
building on previous studies by Hummer \etal
\cite{HummerGarcia1996sjc,hummer1997ion} The development described in
Sec.~\ref{sec:general-theory} casts the resulting finite size
corrections in a different form that facilitates analysis of
interfacial systems. In Sec.~\ref{sec:bulk} we confirm that our
approach is equivalent to that of
Ref.~\onlinecite{HunenbergerMcCammon1999sjc} for a charged solute in
bulk solution. We also verify Hummer's demonstration that simulations,
performed using methods described in Sec.~\ref{sec:SimMethods},
closely follow the predicted approach to the macroscopic limit.

Extended interfaces between coexisting phases can also be simulated
using PBC. The widely used `slab geometry' includes regions of each
phase (e.g., liquid and vapor) that span the periodic simulation cell
in two Cartesian directions ($x$ and $y$), separated by a planar
interface perpendicular to the third direction ($z$).  We show in
Sec.~\ref{sec:SimMethods} that solvation free energies computed from
simulations with this geometry exhibit a strong size dependence that
differs significantly from the bulk case.  Results are presented both
for the SPC/E model of water, and also for a simple polar solvent
comprising diatomic molecules with a pair of opposite charges.  In
Sec.~\ref{sec:slabs} we derive a thermodynamic finite size correction
for ions solvated by an ideal dielectric solvent in the slab geometry.
This correction quantitatively reconciles the size dependence observed
for the simple polar solvent.

For the case of liquid water coexisting with vapor in the slab
geometry, the finite size correction obtained from DCT accounts only
partly for the size dependence observed in simulations.  We attribute
the remaining discrepancy to the intrinsic asymmetry of water's
molecular charge distribution.  In the SPC/E model, positive and
negative charges are not equivalently distributed within each water
molecule, in contrast to the diatomic molecules in our simple polar
solvent. This asymmetry creates thermodynamic biases in the solvation
of cations versus anions, and also underlies the net orientation of
molecules at the air-water interface. In Sec.~\ref{sec:fse-asym} we
show that charge asymmetry in the periodic slab geometry generates a
size-dependent electric potential that can be easily extrapolated to
the macroscopic limit.  The corresponding finite size correction,
added to the dielectric response correction, yields ion solvation free
energies from aqueous slab simulations that are very nearly
independent of unit cell size and aspect ratio. A similar charge
asymmetry correction is applied to results of bulk periodic
simulations in Sec.~\ref{sec:reconciling}, yielding good agreement
with the limiting behavior of slab simulations.

The size dependence of ion solvation free energies near simulated
air-water interfaces thus closely follows predictions of an idealized
continuum model, once effects of charge asymmetry under PBC have been
reckoned. This success of DCT extends to simulation unit cells with
dimensions as small as 1\,nm, indicating a surprising robustness of
linear response theory at nearly molecular length scales. In
Sec.~\ref{sec:discuss} we discuss implications of this result for
microscopic mechanisms and theories of ion solvation near interfaces.
In Sec~\ref{sec:concl} we conclude.


\section{Simulation methods}
\label{sec:SimMethods}

\begin{figure}[tb]
  \includegraphics[width=7.65cm]{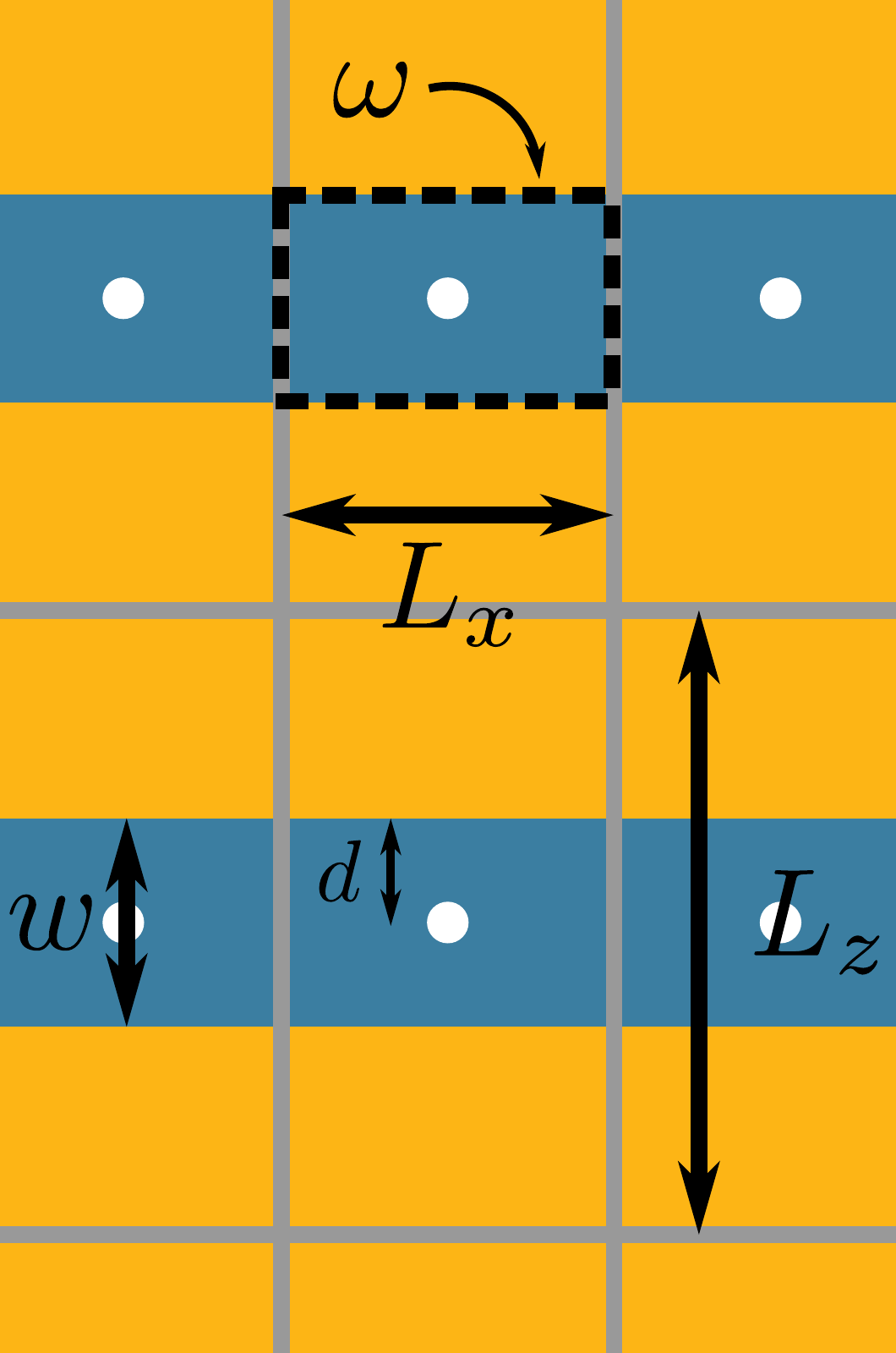}
  \caption{Schematic of the slab geometry typically used in molecular
    simulations of interfacial systems. In the case of a liquid
    coexisting with its vapor, a slab of width $w$ (blue) is separated
    from its periodic replica normal to the interface (taken to be
    along $z$) by a region of vacuum (orange). The length of the
    primary unit cell in the $z$-direction is $L_{z}$. The
    liquid/vapor interface lays in the $xy$-plane, and has a surface
    area $A = L_{x}L_{y}$, where $L_{x}$ and $L_{y}$ are the lengths
    of the simulation cell in $x$ and $y$, respectively. The total
    volume of the simulation cell is $v=L_{x}L_{y}L_{z}$, and the
    volume of the slab per cell is $\omega=Aw$. A solute (white
    circle) is situated in the liquid slab at a distance $d$ from one
    of the liquid/vapor interfaces, and is replicated with the same
    periodicity as the primitive unit cell.}
  \label{fig:slab_schematic}
\end{figure}

The general system of interest, shown schematically in
Fig.~\ref{fig:slab_schematic}, comprises a set of periodically
replicated liquid slabs with width $w$ and dielectric constant
$\epsilon$, separated by regions of vacuum in the direction normal to
the interface, which we take to be along the $z$-direction. The length
of the primary unit cell in the $z$-direction is $L_{\rm z}$. The
liquid/vapor interface lies in the $xy$-plane, and has a surface area
$A = L_{x}L_{y}$, where $L_{x}$ and $L_{y}$ are the lengths of the
simulation cell in $x$ and $y$, respectively. The total volume of the
simulation cell is $v=L_{x}L_{y}L_{z}$. A periodically replicated
solute is situated in the liquid slab at a distance $d$ from one of
the liquid/vapor interfaces. A bulk solution lacking liquid/vapor
interfaces is obtained by setting $w=L_{\rm z}$.

We have numerically examined finite size effects on ion solvation
through molecular simulations of two model solvents. The simpler of
these solvents, diatomic molecules comprising oppositely charged
particles separated by a short distance, closely resembles the
Stockmayer fluid. The other, a simple point charge model of water,
adds the complication of asymmetry under charge inversion, whose
implications for finite size effects are analyzed in
Sec.~\ref{sec:fse-asym}.

Our model solute is the same in both sets of simulations, namely a
Lennard-Jones particle with a point charge at its center.  Parameters
for the Lennard-Jones interaction, $u(r) =
4\varepsilon[(\sigma/r)^{12}-(\sigma/r)^{6}]$ as a function of
distance $r$, are identical to those of the SPC/E model for water,
i.e., $\varepsilon = 0.1553$\,kcal/mol and $\sigma = 3.166$\,\AA. This
choice of $\sigma$ corresponds roughly to the size of a Ca$^{2+}$ ion
\cite{HummerGarcia1996sjc}. (The leading order correction considered
in this work is, however, independent of solute size.)  For
consistency with standard notation, we use the symbol $\sigma$ here to
denote the Lennard-Jones diameter. The same symbol will later denote a
different quantity, namely the surface charge that results from
dielectric response. The distinction should be clear from context.

In water studies, we employed the SPC/E model for interactions
involving solvent molecules\cite{BerendsenStraatsma1987sjc}.  This
choice, made for simplicity, affects the exact values of computed
solvation free energies, but our conclusions about finite size effects
should apply equally well to more elaborate molecular models. Bulk
liquid water simulations included a single solute and a number $N$ of
water molecules ranging from 64 to 756, such that the total number
density was $0.03333$\,\AA$^{-3}$.  Simulations of air-water
coexistence included one solute and $N=128$ water molecules in a
simulation cell with lateral dimensions $L_{x} = L_{y} =
12.429$\,\AA{} and varying separation between periodic liquid slabs.
(As the sole exception, in Fig.~\ref{fig:bulk_comparison} we show
results for $L_x=L_y=31.1$\,\AA{} as well.) The resulting aspect
ratios $L_z/L_x$ ranged from 4 to 30. To suppress interfacial
instabilities at very high aspect ratios, we introduced an external
confining potential, specifically walls that effectively constrain
molecules to a region slightly larger than the neat liquid slab
occupies.  These walls interact with the oxygen atom of each water
molecule through a potential $u_{\text{solv-wall}}(z)$, which inside
the walls ($|z| < |z_{\text{wall}}|$) has a Weeks-Chandler-Anderson
form \cite{WCA}: $u_{\text{solv-wall}}(z) =
4\varepsilon_{\text{solv-wall}}[(\delta
  z/\sigma_{\text{solv-wall}})^{-12} - (\delta
  z/\sigma_{\text{solv-wall}})^{-6} + 1/4]$ for $|\delta z| <
2^{1/6}\sigma_{\text{solv-wall}}$, where $\delta z = z -
z_{\text{wall}}$, and $z_{\text{wall}} = \pm 15$\,\AA{}.  Outside the
walls ($|z|>|z_{\text{wall}}|$), $u_{\text{solv-wall}}(z)$ is
infinite. Otherwise ($|\delta z| > 2^{1/6}\sigma_{\text{solv-wall}}$),
$u_{\text{solv-wall}}(z)$ vanishes. The coordinate $-L_{z}/2 < z <
L_{z}/2$ is the molecule's position within the primary simulation
cell. Away from the interfaces, the density of the liquid varied
approximately between 0.03200\,\AA$^{-3}$ and 0.03333\,\AA$^{-3}$,
depending on the aspect ratio and the charge of the solute.

The simpler polar fluid we consider consists of `dumbbell' molecules,
each a pair of Lennard-Jones spheres ($\varepsilon = 0.1553$\,kcal/mol
and $\sigma = 3.166$\,\AA) separated by a rigid bond of length
$l_{\text{D}} = 0.25$\,\AA. Opposite charges of magnitude
$|q_{\text{D}}| = 0.9\,e$ reside at the centers of these spheres.  We
consider a liquid state density of $0.025$\,\AA$^{-3}$.  Bulk
simulations included a number $N$ of dumbbells ranging from 48 to
384. Slab simulations included $N=96$ molecules. As in aqueous
simulations, lateral dimensions were fixed at $L_{x} = L_{y} =
12.429$\,\AA, while the aspect ratio $L_z/L_x$ varied from 4 to
30. Walls were again used to confine solvent molecules to the range
$-15.0\,\text{\AA} < z <15.0\,\text{\AA}$. However, in this case, the
confining potential interacts with both Lennard-Jones spheres of each
dumbbell molecule.

Some of the finite size corrections detailed below involve the static
dielectric constant $\epsilon$. In the case of SPC/E water the value
of $\epsilon$ has been established by previous work, and is
sufficiently large that required factors of $(\epsilon-1)/\epsilon$
are nearly unity.  The dumbbell fluid is considerably less
polarizable, so that a more precise estimate of $\epsilon$ is needed.
From a 5\,ns bulk simulation of 512 dumbbell molecules (and no
solute), we estimated the dielectric constant to be $\epsilon \approx
7.1$, obtained from the fluctuation-dissipation theorem:
\cite{neumann1983dipole}
\begin{equation}
  \epsilon -1 =
  \frac{4\pi\beta}{3V}\left(\langle\mbf{M}^{2}\rangle_{\mbf{E}=\mbf{0}}
  - \langle\mbf{M}\rangle^{2}_{\mbf{E}=\mbf{0}}\right)
\end{equation}
\noindent where $\mbf{M}$ is the total dipole moment of the simulation
cell, and $\beta = (k_{\text{B}}T)^{-1}$ ($k_{\text{B}}$ is
Boltzmann's constant). We append the subscript `$\mbf{E}=\mbf{0}$' to
emphasize that this formula holds with the use of tin foil boundary
conditions (see
e.g. Refs.~\onlinecite{zhang2016computing,de1986computer,neumann1983dipole}). Similar
dumbbell models have been examined in previous
studies\cite{reif2016origin}, which reported similar values of
$\epsilon$ for comparable molecular dipole moments.

All simulations were performed with the \textsmaller{LAMMPS}
simulation package \cite{plimpton1995sjc}. Dynamics were propagated at
constant volume $V$ and temperature $T=298$\,K using Langevin dynamics
as implemented in \textsmaller{LAMMPS}
\cite{SchneiderStoll1978sjc,DunwegWolfgang1991sjc}, with a time step
of 1\,fs. The \textsmaller{SHAKE} algorithm was used to constrain all
bonds and angles \cite{ryckaert1977numerical,andersen1983rattle}. Long
range electrostatic interactions were evaluated using Ewald summation
with tin foil boundary conditions.  (The solvent's net dipole would
experience different forces under a different choice of boundary
conditions\cite{LPS1,LPS2,smith1981electrostatic,redlack1975coulombic,neumann1983dipole,kantorovich1999coulomb,ballenegger2014communication}.
The solvent response of interest, however, does not directly involve
this dipole, so we expect the choice of boundary conditions to have
minor significance.) Ewald sums were calculated using the particle-particle
particle-mesh solver \cite{HockneyEastwood1988sjc}, with parameters
chosen such that the RMS error in the forces were a factor $10^{5}$
smaller than the force between two unit charges separated by a
distance of 1.0\,\AA \cite{kolafa1992cutoff}.

The solvation free energies $F_{\rm chg}^{(\mbf{L})}(q)$ we consider are
precisely defined by
\begin{widetext}
  \begin{equation}
    \label{eqn:expFchg}
    \exp[-\beta F_{\rm chg}^{(\mbf{L})}(q)] =
    \langle \exp[-\beta q \phi_{\rm solv}]\rangle_0^{(\mbf{L})} =
    \int\!\mrm{d}\phi_{\rm solv}\,P^{(\mbf{L})}(\phi_{\rm solv};0) \exp[-\beta q \phi_{\rm solv}]
  \end{equation}
\end{widetext}
where $\phi_{\rm solv}$ is the electric potential generated by the
solvent, evaluated at the center of a solute; $P^{(\mbf{L})}(\phi_{\rm
  solv};q)$ is its probability distribution in the presence of a
solute charge $q$; and $\langle \cdot \rangle_q^{(\mbf{L})}$ denotes
an average over $P^{(\mbf{L})}(\phi_{\rm solv};q)$. The superscript
$({\bf L})$ specifies the dimensions $\mbf{L} = (L_{x},L_{y},L_{z})$
of the simulated unit cell.  In the limit $\mbf{L}\to\infty$ of large
simulation cell size, $F_{\rm chg}^{(\mbf{L})}(q)$ approaches the
reversible work of charging a solute at infinite dilution.  Evaluating
the integral in Eq.~\ref{eqn:expFchg}, which is a statement of the
potential distribution theorem \cite{widom1982potential}, requires
knowledge of $P^{(\mbf{L})}(\phi_{\rm solv};0)$ in its extreme wings,
which we obtained from umbrella sampling. Specifically, we introduced
a series of solute charges $q/e=-1.0,-0.9,\ldots,0,\ldots,+0.9,+1.0$,
which bias typical solvent potential fluctuations to a corresponding
series of ranges centered on $\bar{\phi}_{\rm solv}(q)$, where
\begin{equation}
\frac{d \ln P^{(\mbf{L})}(\phi_{\rm
    solv};0)}{d\phi_{\rm solv}}
\bigg|_{\phi_{\rm solv}=\bar{\phi}_{\rm solv}} = \beta q
.
\end{equation}
Samples from the probability distributions $P^{(\mbf{L})}(\phi_{\rm
  solv};q) = P^{(\mbf{L})}(\phi_{\rm solv};0) e^{-\beta q \phi_{\rm
    solv}}/\langle e^{-\beta q \phi_{\rm solv}}\rangle_0$ were then
reweighted and combined according to the \textsmaller{MBAR} algorithm
\cite{ShirtsChodera2008sjc}. The resulting statistics of extreme
solvent fluctuations required to compute $F^{(\bf L)}_{\rm chg}(q)$
could as well be obtained from alternative methods such as
thermodynamic integration, reversible work calculation, or
irreversible transformations via Jarzynski's identity
\cite{HummerGarcia1996sjc,darden1998ionic,jarzynski1997nonequilibrium}. The
biasing method we used is especially convenient for molecular dynamics
simulation, as it can be achieved simply by modulating the charge of a
single particle.

The value of $\phi_{\rm solv}(\mbf{R}^{N})$ for a given configuration
$\mbf{R}^{N}$ of solvent molecules was evaluated by inserting a solute
test charge, and then subtracting contributions due to the test charge
itself. In detail, we compute the total potential energy
$u(\mbf{R}^{N};q_{\rm test})$ of a system including solvent, a
periodic collection of test charges $q_{\rm test}$, and a charge
neutralizing background. From this we first subtract the energy
$u(\mbf{R}^{N};0)$ of interactions among solvent molecules and of
Lennard-Jones interactions with the solute.  We then subtract the
energy of interactions among periodic test charges and the
neutralizing background, $q_{\rm test} \phi_{\rm wig}/2$, where
$\phi_{\rm wig}$ is the canonical Wigner potential discussed in
Sec.~\ref{sec:bulk} and Appendix~\ref{sec:generalized-Wigner}.
Dividing the remainder by $q_{\rm test}$ yields the solvent potential,
$\phi_{\rm solv}(\mbf{R}^{N}) = q_{\rm
  test}^{-1}[u(\mbf{R}^{N};q_{\rm test})-
  u(\mbf{R}^{N};0)-q_{\rm test} \phi_{\rm wig}/2]$.

Charging free energies $F_{\rm chg}^{(\mbf{L})}(q)$ computed with
these methods are presented in Figs.~\ref{fig:bulk_results}
and~\ref{fig:slab_results}, for bulk and interfacial systems,
respectively. A strong dependence on periodicity $\mbf{L}$ is evident
for both model solvents considered. For cubic simulation cells bulk
free energies vary by 10s of $k_{\rm B}T$ as cell dimensions grow from
roughly 1 to 2~nm. Bulk results for anisotropic cells differ even more
dramatically, by 100s of $k_{\rm B}T$ as the aspect ratio $L_z/L_x$
grows from 2 to 12, and appear not to converge as $L_z\to\infty$ with
fixed $L_x$. Similarly large finite size effects are exhibited by
interfacial simulations. In this case, however, charging free energies
do appear to converge in the limit of infinite aspect ratio, pointing
to qualitative differences in the system size dependence of periodic
bulk and slab geometries. Such nontrivial convergence behavior also
warns that extrapolating results naively from ever-larger systems may
not achieve physically meaningfully asymptotic properties.


\section{Correcting for finite size effects on ionic solvation free energies}
\label{sec:general-theory}

This work is by no means the first to address the role of finite size
effects in ionic solvation free energies obtained from molecular
simulation
\cite{HummerGarcia1996sjc,figueirido1997finite,hummer1997ion,HunenbergerMcCammon1999sjc,HummerGarcia1998sjc},
and indeed, the central assumption---that DCT adequately describes
long-wavelength dielectric response---follows the perspective of
Ref.~\onlinecite{HunenbergerMcCammon1999sjc}. The majority of previous
work that address finite size effects has, however, focused on a
scenario that lacks macroscopic dielectric boundaries, such as the
liquid/vapor interface. Here we present a general formalism that is
also applicable in such cases.

The quantity that we seek to calculate is the free energy to introduce
a charge $q$ to the center of a neutral cavity embedded in a solvent
with dielectric constant $\epsilon$, at infinite dilution:
\begin{equation}
  F_{\rm chg}^{(\infty)}(q) \equiv F^{(\infty)}(q) - F^{(\infty)}(0)
\end{equation}
\noindent
$F^{(\infty)}(q)$ is the free energy of a macroscopic system in the
presence of a single solute with charge $q$.  Molecular simulations
cannot directly access $F_{\rm chg}^{(\infty)}(q)$, but can instead
provide corresponding free energies under PBC:
\begin{equation}
  F_{\rm chg}^{(\mbf{L})}(q) \equiv F^{(\mbf{L})}(q) - F^{(\mbf{L})}(0)
\end{equation}
As in Sec.~\ref{sec:SimMethods} the superscript $({\bf L})$ indicates
spatial periodicity $\mbf{L} = (L_{x},L_{y},L_{z})$ in three Cartesian
directions.  Our goal is to connect the charging free energies for
periodic and macroscopic systems, i.e., to calculate a correction
\begin{equation}
\Delta F({\bf L};q) = F_{\rm chg}^{(\infty)}(q) -
  F_{\rm chg}^{(\mbf{L})}(q) 
\end{equation}
that can be applied to simulation results as an extrapolation
to infinite dilution.

We consider two distinct contributions to $\Delta F({\bf L};q)$.  The
first accounts for size dependence of a solvent's structural response
to solute charging, which we estimate using dielectric continuum
theory. The second considers biases on ion solvation that exist
even in the unperturbed solvent, specifically electric potentials
experienced by a neutral solute, which are also size-dependent.

\subsection{Size-dependent dielectric response}
DCT is defined by a simple linear response relationship between the
electric field at a point ${\bf r}$ and the average induced
polarization density $\mbf{m}(\mbf{r})$ at that location. Together with
Poisson's equation, this relation connects spatial variations in
$\mbf{m}(\mbf{r})$ to the density $\rho(\mbf{r})$ of free charge,
\begin{equation}
  \label{eqn:DCT}
  \nabla\cdot\mbf{m} = \bigg(\frac{\epsilon-1}{\epsilon}\bigg)
  \rho(\mbf{r})
\end{equation}
(Throughout this formulation, we work in a unit system in which
$4\pi\epsilon_{0} = 1$, where $\epsilon_{0}$ is the permittivity of
free space.) In our case the free charge represents periodically
arranged solutes and their compensating uniform background of charge
density $\bar{\rho} = q/(L_{x}L_{y}L_{z})$,
\begin{equation}
  \label{eqn:rho_solute}
  \rho(\mbf{r}) = \sum_{\mbf{b}} \rho_{0}(\mbf{r}+\mbf{b}) - \bar{\rho}
\end{equation}
\noindent Here, $\rho_{0}(\mbf{r})$ is the charge distribution of a
single solute, localized at position $\mbf{r}_0$ inside the dielectric
and with net charge $q = \int \!\mrm{d}\mbf{r}\,\rho_{0}(\mbf{r})$.
The solute's periodic replicas are separated by lattice vectors
$\mbf{b} = (n_{x}L_{x},n_{y}L_{y},n_{z}L_{z})$, with $n_{x}$, $n_{y}$
and $n_{z}$ taking on all integer values.  If dielectric boundaries
are present, we take them to have the same spatial periodicity as the
solutes, as well as a clear dependence on $\mbf{L}$. For the periodic
slab geometry these boundaries are a pair of infinite parallel planes
with fixed separation $w$, repeated in the $z$ direction with
periodicity $L_z$ (see Fig.~\ref{fig:slab_schematic}).

To calculate the charging free energy $F_{\rm chg}^{(\mbf{L})}(q)$
for this theory, we exploit a generic property of linear response,
namely that the change in free energy due to an external field
is half the average interaction energy between system and field.
For a dielectric material interacting with periodic solutes and
their neutralizing background, 
\begin{equation}
  \label{eqn:DeltaFL}
  F_{\rm chg}^{(\mbf{L})}(q) =
  \frac{1}{2} q \langle\phi_{\text{solv}}(\mbf{r}_{0})\rangle_{q}^{(\mbf{L})},
\end{equation}
where $\phi_{\text{solv}}(\mbf{r})$ is the electric potential
generated by the polarization field $\mbf{m}(\mbf{r})$. Because the
solvent is electricially neutral, its interaction with the uniform
background charge does not contribute to Eq.~\ref{eqn:DeltaFL}.
Determining a finite size correction $\Delta F_{\rm DCT}({\bf L};q)$
from dielectric continuum theory thus amounts to calculating the
average solvent potential at the center of a solute.

We obtain the solvent potential by integrating over the volume
$\Omega$ occupied by the solvent:
\begin{equation}
\langle\phi_{\text{solv}}(\mbf{r}_{0})\rangle_{q}^{(\mbf{L})}
= \int_{\Omega}\!\mrm{d}\mbf{r}\,\mbf{m}(\mbf{r})\cdot\nabla
\frac{1}{|\mbf{r}-\mbf{r}_0|}
\end{equation}
This volume is periodic by construction but it need not be
connected. If $\Omega$ comprises disconnected regions, then a periodic
series of surfaces $\partial\Omega$ bounds these regions. The unit
normal vector on these boundaries we denote by
$\hat{\mbf{n}}$. Integrating by parts and using the divergence theorem
gives:
\begin{equation}
  \langle\phi_{\text{solv}}(\mbf{r}_{0})\rangle_{q}^{(\mbf{L})}=
  \int_{\partial\Omega}\!\mrm{d}\mbf{R}\,\frac{\hat{\mbf{n}}\cdot\mbf{m}
    (\mbf{R})}{|\mbf{R}-\mbf{r}_0|} -
  \int_{\Omega}\!\mrm{d}\mbf{r}\,\frac{\nabla\cdot\mbf{m}(\mbf{r})}
      {|\mbf{r}-\mbf{r}_0|}
\end{equation}
where $\mbf{R}$ is a point on the set of boundaries $\partial\Omega$.
The normal component of solvent
polarization at a boundary, 
$\sigma(\mbf{R};\mbf{L}) \equiv \hat{\mbf{n}}\cdot\mbf{m} (\mbf{R})$,
serves as an effective surface charge density due to polarization.
Using the basic linear response relation in Eq.~\ref{eqn:DCT},
we finally obtain:
\begin{equation}
  \label{eqn:phisolv}
  \langle\phi_{\text{solv}}(\mbf{r}_{0})\rangle_{q}^{(\mbf{L})}=
  \int_{\partial\Omega}\!\mrm{d}\mbf{R}\,\frac{\sigma(\mbf{R};\mbf{L})}
      {|\mbf{R}-\mbf{r}_0|} -
      \dctfac\int_{\Omega}\!\mrm{d}\mbf{r}\,\frac{\rho(\mbf{r})}
                 {|\mbf{r}-\mbf{r}_0|}
\end{equation}

The correction $\Delta F_{\rm DCT}({\bf L};q)$ compares response at
finite $\mbf{L}$ with the macroscopic result.  In the limit
$\mbf{L}\to\infty$, all solute replicas described by $\rho(\mbf{r})$
are irrelevant except for the central charge at $\mbf{r}_0$, and the
background charge density vanishes. As a result,
\begin{widetext}
  \begin{equation}
    \label{eqn:Deltaphisolv_general}
    \Delta F_{\rm DCT}({\bf L};q) = \frac{q}{2}\bigg[
      -\int_{\partial\Omega}\!\mrm{d}\mbf{R}\,\frac{\Delta\sigma(\mbf{R})}
      {|\mbf{R}-\mbf{r}_0|}
      + q\dctfac\int_{\Omega}\!\mrm{d}\mbf{r}\,
      \frac{1}{|\mbf{r}-\mbf{r}_0|}\bigg(\sum_{\mbf{b}\neq\mbf{0}}
      \delta(\mbf{r}-\mbf{r}_0+\mbf{b}) - (L_{x}L_{y}L_{z})^{-1}\bigg)\bigg],
  \end{equation}
\end{widetext}
\noindent where $\Delta\sigma(\mbf{R}) = \sigma(\mbf{R};\mbf{L}) -
\sigma(\mbf{R};\infty)$ is a difference between periodic and
macroscopic systems. Motivated by the slab geometry of interest, we
have assumed that there is a clear correspondence between boundaries
for finite and infinite $\mbf{L}$.  The $\mbf{L}\to\infty$ slab
possesses just two planar surfaces (neglecting the solute's excluded
volume\footnote{Acknowledging that the dielectric does not
    penetrate the solute is essential for local polarization response,
    which would otherwise be singular. This local response, however,
    should be consistent across different system sizes; a comparison
    among them eliminates the singularity and justifies treating the
    solute as a point charge within the dielectric domain
    $\Omega$.}), bounding the
solvent from above and below; these two surfaces are also present at
finite $\mbf{L}$. On all other boundaries included in
$\partial\Omega$, there is no macroscopic contribution, i.e.,
$\sigma(\mbf{R};\infty)=0$. A more complicated set of periodic
boundaries might not permit this simplification, requiring instead an
explicit subtraction of surface integrals evaluated on different sets
of boundaries.

For any choice of boundaries, evaluating
Eq.~\ref{eqn:Deltaphisolv_general} requires knowledge of the function
$\sigma(\mbf{R};\mbf{L})$. We have not solved a general dielectric
boundary value problem, but have instead rephrased it in a convenient
way. For the slab geometry, $\sigma(\mbf{R};\mbf{L})$ can be worked
out exactly in terms of a series of image charges, as detailed in
Appendix~\ref{sec:image_charges}. More importantly,
$\Delta\sigma(\mbf{R})$ is amenable to a greatly simplifying
approximation, which we will show to be very accurate in the case of
periodic slabs. For a bulk system lacking boundaries,
Eq.~\ref{eqn:Deltaphisolv_general} does constitute a full solution.
In the next section we show that this result is consistent with
previous work on ion solvation in bulk solvents.


\subsection{Application to ion solvation in the absence of interfaces}
\label{sec:bulk}

Understanding the behavior of $\Delta F(\mbf{L};q)$ for `bulk'
periodic systems -- those lacking any macroscopic dielectric boundary --
has been the focus of much research. In an early study on the subject,
Hummer \etal{} \cite{HummerGarcia1996sjc} argued that estimates of
charging free energy should include interactions of the ion with its
periodic replicas and background charge. For simulations of liquid
water using Ewald summation and a cubic cell of side length $L$, they
showed that incorporating these `self-interactions' can yield free
energies that are essentially independent of system size. The finite
size correction is thus well approximated in this case by
\begin{align}
  \label{eqn:HPG_infinite-eps}
  \Delta F(\mbf{L};q) \approx \frac{1}{2}q\phi_{\text{wig}}
\end{align}
\noindent
where $\phi_{\text{wig}}/q \approx -2.837297/L$ (see
e.g. Ref.~\onlinecite{NijboerRuijgrok1988sjc}, and note that atomic
units are implied in the numerical value).  The Wigner potential
$\phi_{\text{wig}}/q$ is defined as the electric potential at the site
of a unit point charge due to all of its periodic replicas and a
homogeneous background charge that acts to neutralize the primitive
cell. ($\phi_{\text{wig}}/q$ is commonly referred to as
$\xi_{\text{EW}}$ in the literature.)  Based on a dimensional
analysis, Figueirido \etal{} \cite{figueirido1997finite} proposed the
following correction for finite dielectric constant $\epsilon$:
\begin{equation}
  \label{eqn:HPG_finite-eps}
  \Delta F(\mbf{L};q)\approx \frac{1}{2}q\dctfac\phi_{\text{wig}}
\end{equation}
\noindent which has the correct limiting behavior for $\epsilon\to
1$. Later works extended these corrections to also account for the
finite size of the ion
\cite{hummer1997ion,HunenbergerMcCammon1999sjc}. We do not consider
such higher order corrections in this article.

\begin{figure*}[tb]
  \includegraphics[width=16.02cm]{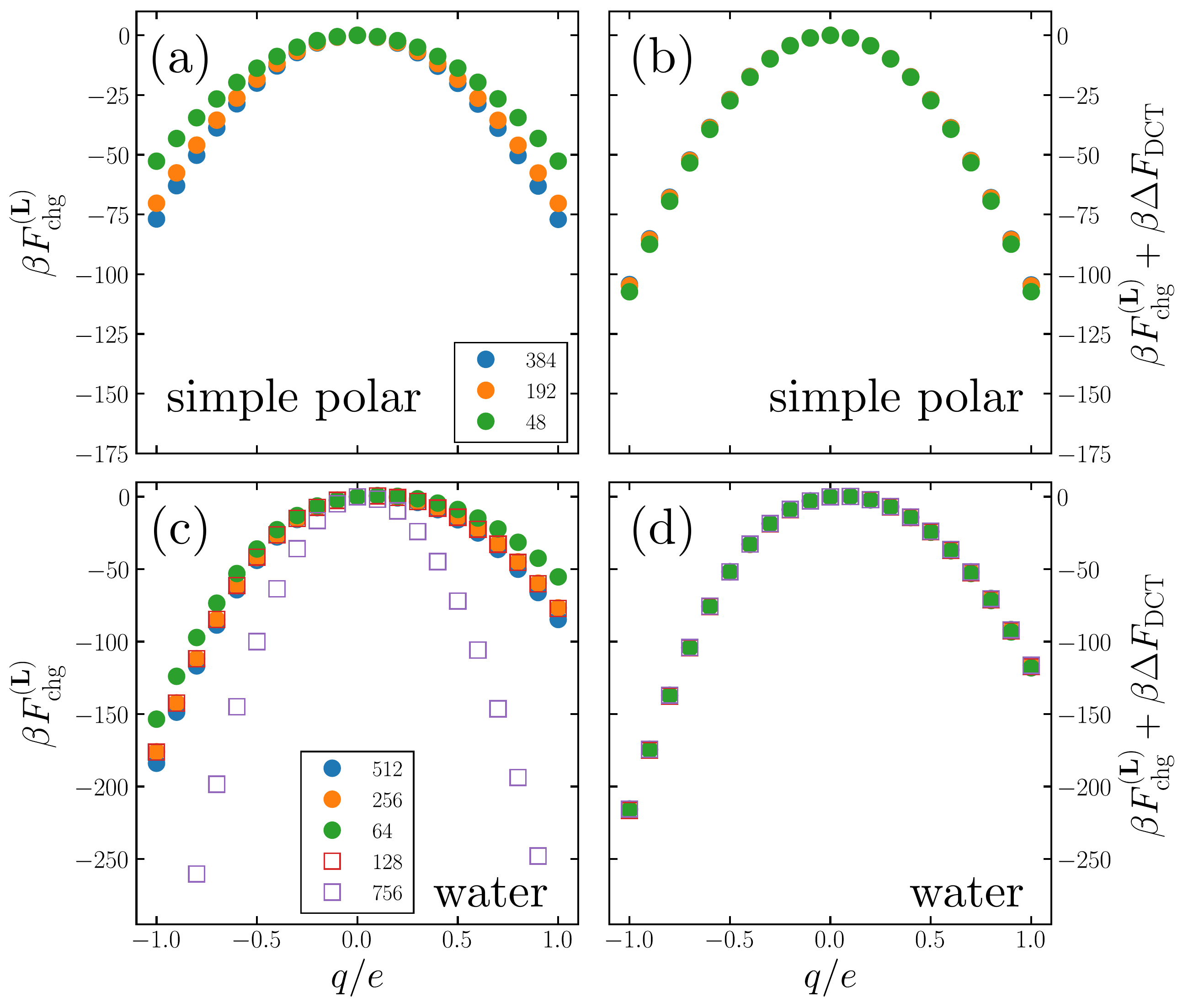}
  \caption{System size dependence of solute charging free energy
    $F_{\text{chg}}^{(\mbf{L})}(q)$, as a function of solute charge
    $q$, for periodic {\em bulk} simulations.  (a) Results for the
    simple polar fluid, with 48, 192, and 384 molecules (as indicated
    in the legend box) exhibit strong finite size effects. These
    variations with cell size are largely reconciled in (b) by the
    dielectric correction $\Delta F_{\text{DCT}}$ from
    Eq.~\ref{eqn:HPG_finite-eps}. Higher-order corrections due to the
    finite size of the solute
    \cite{hummer1997ion,HunenbergerMcCammon1999sjc} have been
    neglected.  (c) Results for SPC/E water also vary substantially
    with cell size, and also with its aspect ratio.  The filled
    circles were obtained with 64, 256 and 512 molecules in cubic
    simulation cells, while the empty squares were obtained with 128
    and 756 molecules in cuboidal cells ($L_{z}/L_{x} = 2$ and $12$,
    respectively), as indicated in the legend box. Note that the
    results for 756 molecules extend far beyond the data range shown
    ($\beta F_{\text{chg}}^{(\mbf{L})}(-e) \approx -408$).  (d)
    Corrected free energies $F_{\text{chg}}^{(\mbf{L})}(q) + \Delta
    F_{\text{DCT}}(\mbf{L};q)$ for SPC/E water (obtained using
    Eqs.~\ref{eqn:HPG_finite-eps} and~\ref{eqn:canonical-wigner} for
    cubic and cuboidal simulation cells, respectively) are nearly
    indistinguishable on this energy scale.}
  \label{fig:bulk_results}
\end{figure*}

In the absence of interfaces, our DCT expression in 
Eq.~\ref{eqn:Deltaphisolv_general}
simplifies to
\begin{widetext}
  \begin{equation}
    \label{eqn:canonical-wigner}
    \Delta F_{\rm DCT}(\mbf{L};q) = \frac{1}{2}q^2 \dctfac
    \int_{\Omega}\!\mrm{d}\mbf{r}\,\frac{1}{r}\bigg[\sum_{\mbf{b}\neq\mbf{0}}\delta(\mbf{r}+\mbf{b}) - (L_{x}L_{y}L_{z})^{-1}\bigg]
  \end{equation}
\end{widetext}
For a cubic primitive cell ($L_x=L_y=L_z=L$), this result is
equivalent to Figueirido's, as expected from the DCT analysis of
Ref.~\onlinecite{HunenbergerMcCammon1999sjc}. Fig.~\ref{fig:bulk_results}
exemplifies the remarkable effectiveness of this correction,
echoing the conclusions of Hummer \etal{}  Here we have added $\Delta
F_{\rm DCT}(\mbf{L};q)$ to the values of $F_{\rm chg}^{(\mbf{L})}(q)$
obtained from bulk molecular simulations of water (d) and the simple
polar fluid (b).
The DCT correction reconciles solvation free
energies from simulations with significantly different periodicity,
reducing discrepancies of $\sim 25 k_{\rm B}T$ down to $\sim 1 k_{\rm
  B}T$. Residual finite size effects beyond $\Delta F_{\rm
  DCT}(\mbf{L};q)$ are noticeable only for the smallest simulations of
the simple polar liquid. We have confirmed numerically that these
differences can be largely removed with higher-order DCT corrections
that account for the solute's excluded volume
\cite{hummer1997ion,HunenbergerMcCammon1999sjc}.

For an anisotropic cell, Eq.~\ref{eqn:canonical-wigner} is a very
straightforward generalization of the standard correction. To evaluate
it numerically, we have used Ewald summation as outlined in
Appendix~\ref{sec:generalized-Wigner}. In Fig.~\ref{fig:bulk_results}
we include simulation results for cuboidal simulation cells with
$L_{z}/L_{x} = 2$ and $12$, demonstrating that the DCT correction
remains accurate even as the cell's aspect ratio becomes large. The
interesting aspect of this comparison is that $\Delta F_{\rm
  DCT}(\mbf{L};q)$ diverges as $L_{z}/L_{x} \to\infty$ with $L_x L_y$
fixed. The accuracy of our correction then suggests that $F_{\rm
  chg}^{(\mbf{L})}(q)$ diverges in the opposite sense with growing
aspect ratio. Indeed, simulation results with 756 molecules and
$L_{z}/L_{x} = 12$ extend far beyond the plotted energy scale ($\beta
F_{\text{chg}}^{(\mbf{L})}(-e) \approx -408$), hinting at such a
divergence.

\subsection{Application to periodic slabs}
\label{sec:slabs}

Applying the general result in Eq.~\ref{eqn:Deltaphisolv_general} to
systems with boundaries is considerably more difficult, as it requires
calculation of the polarization surface charge
$\sigma(\mbf{R};\mbf{L})$ (as well as integration of
$\rho(\mbf{r})/|\mbf{r}-\mbf{r}_0|$ over disconnected volumes). The
spatial variation of $\sigma(\mbf{R};\mbf{L})$ encodes important
driving forces of solvation. A point ion near a dielectric boundary
experiences an electric field that grows sharply in magnitude as it
approaches the surface.  For a semi-infinite dielectric with a planar
boundary, this field is equivalent in DCT to the force generated by an
image charge opposite the boundary, accompanied by strong spatial
variation in polarization surface charge.  A poor description of this
position dependence would yield a poor approximation to
$\langle\phi_{\text{solv}}(\mbf{r}_{0})\rangle_{q}^{(\mbf{L})}$.  This
feature of $\sigma(\mbf{R};\mbf{L})$, however, is predominantly local
in space and therefore only weakly sensitive to PBC.  Moreover, such
local aspects of solvation are likely not well described by DCT, which
lacks microscopic complexities that can dominate on small scales.  The
{\em difference} quantity $\Delta\sigma(\mbf{R};\mbf{L})$, we argue,
should have much weaker spatial dependence, and should be well
described by DCT. 

For the specific case of periodic slabs, a solution for
$\sigma(\mbf{R};\mbf{L})$ can be obtained by summing contributions
from an infinite series of effective image charges, as described in
Appendix~\ref{sec:image_charges}.  This solution exhibits the rapid
spatial variation described above, particularly as the solute ion
nears the dielectric interface.  But as anticipated, the image charges
that contribute most strongly to this position dependence are
identical in value and placement when $\mbf{L}\to\infty$. The finite
size effects of interest to our work originate instead in the
long-wavelength features of polarization surface charge. We thus
propose a greatly simplifying approximation, namely that
$\Delta\sigma(\mbf{R}) \approx \bar{\sigma}$ is constant along the
planar boundaries of each dielectric slab.  (This argument relies on
the ion residing within the solvent's liquid phase. As discussed in
Appendix~\ref{sec:image_charges}, the approximation is inappropriate
when $\mbf{r}_0$ lies in the vapor phase instead.)

An appropriate value for $\bar{\sigma}$ can be determined by
integrating over a single periodic replica $\partial\omega$ of the
slab's boundaries:\footnote{We take the net surface charge on the
  slab's two boundaries to be equal. For a single aperiodic slab, this
  symmetry holds regardless of the ion's position in the slab, as can
  be shown by summing image charges above and below each
  interface. For the periodic case, we assume that symmetrically
  placed replicas do not upset this balance
  (Figs.~\ref{fig:slab_images} and \ref{fig:slab_images_per}).}
\begin{equation}
  2 L_x L_y \bar{\sigma} =
  \int_{\partial\omega}\!\mrm{d}\mbf{R}\,\hat{\mbf{n}}\cdot\mbf{m}(\mbf{R})
  = \int_{\omega}\!\mrm{d}\mbf{r}\,\nabla\cdot\mbf{m}(\mbf{r}),
\end{equation}
where $\omega$ denotes a single periodic replica of the slab (see
Fig.~\ref{fig:slab_schematic}).  In using the divergence theorem, we
have exploited the fact that symmetry and continuity of
$\mbf{m}(\mbf{r})$ within the slab requires that
$\hat{\mbf{n}}\cdot\mbf{m}(\mbf{R})=0$ at the lateral interfaces
between periodic replicas. Use of Eqs.~\ref{eqn:DCT}
and~\ref{eqn:rho_solute} then yields
\begin{equation}
  \label{eqn:sigma_bar}
  \bar{\sigma} = \frac{q}{2A}\dctfac\bigg[1 - \frac{w}{L_{z}}\bigg],
\end{equation}
where $w$ is the slab's thickness and $A=L_x L_y$ is the area of a
single periodic replica of the upper (or lower) dielectric boundary.
This value of $\bar{\sigma}$ is precisely what is required for $\Delta
F_{\rm DCT}(\mbf{L};q)$ to be finite.

The finite size correction $\Delta F_{\rm DCT}^{({\rm
    unif})}(\mbf{L};q)$ that results from this uniform surface charge
approximation, together with Eq.~\ref{eqn:sigma_bar},
\begin{widetext}
\begin{eqnarray}
  \label{eqn:DeltaFunif}
  \Delta F_{\rm DCT}^{({\rm unif})}(\mbf{L};q) =
  &&\frac{q^2}{2}\dctfac\bigg[
    -\frac{1}{2A}
    \bigg(1 - \frac{w}{L_{z}}\bigg)
    \int_{\partial\Omega}\!\mrm{d}\mbf{R}\,\frac{1}
        {|\mbf{R}-\mbf{r}_0|}
        \nonumber
        \\ 
      &+&\int_{\Omega}\!\mrm{d}\mbf{r}\,
      \frac{1}{|\mbf{r}-\mbf{r}_0|}\bigg(\sum_{\mbf{b}\neq\mbf{0}}
      \delta(\mbf{r}-\mbf{r}_0+\mbf{b}) - (L_{x}L_{y}L_{z})^{-1}\bigg)\bigg],
\end{eqnarray}
\end{widetext}
has a simple physical interpretation. The quantity in square brackets
in Eq.~\ref{eqn:DeltaFunif} corresponds to an electric potential
$\phi_{\rm wig}^*$ generated by: (i) periodically arranged unit point
charges (excepting the central replica); (ii) a partially neutralizing
charge density that is uniform within the dielectric slabs and
vanishes outside; and (iii) uniformly charged plates at the dielectric
boundaries, which achieve overall neutralization.  It can be regarded
as a generalization of the Wigner potential, in which the uniform
compensating charge outside the dielectric has been moved to the
dielectric boundaries. Interestingly, in the limit $L_{z}\to\infty$
with $A$ and $w$ fixed, all of the compensating charge moves to these
boundaries. This scenario -- a single dielectric slab, with ions
periodically replicated in the $x$ and $y$ directions, bounded above
and below by uniformly charged neutralizing plates -- is markedly
similar to the neutralization scheme proposed by Ballenegger \etal{}
\cite{BalleneggerCerda2009sjc} for molecular dynamics simulations of
charged slab systems that are aperiodic in the $z$ direction.

The modification of the Wigner potential described above can be
calculated exactly in closed form, as shown in
Appendix~\ref{sec:generalized-Wigner}. The DCT finite size
correction can then be written simply as
\begin{equation}
  \label{eqn:DeltaFunif-closedform}
  \Delta F_{\rm DCT}^{\rm (unif)} = \frac{1}{2}q\dctfac
  \bigg[\phi_{\text{wig}} - \frac{\pi q}{3 A L_z^2}(L_z-w)^3
    \bigg],
\end{equation}
In the limit that $L_x$, $L_y$, and $L_z$ all become infinite, with
$w$ still fixed, $\Delta F_{\rm DCT}^{({\rm unif})}(\mbf{L};q)\to 0$,
an important property of the full solution for $\Delta F_{\rm DCT}$
that is preserved by the uniform surface charge approximation. Note
that $\Delta F_{\rm DCT}^{\rm (unif)}$ does not depend upon the
solute's position within the slab.

\begin{figure*}[tb]
  \includegraphics[width=16.02cm]{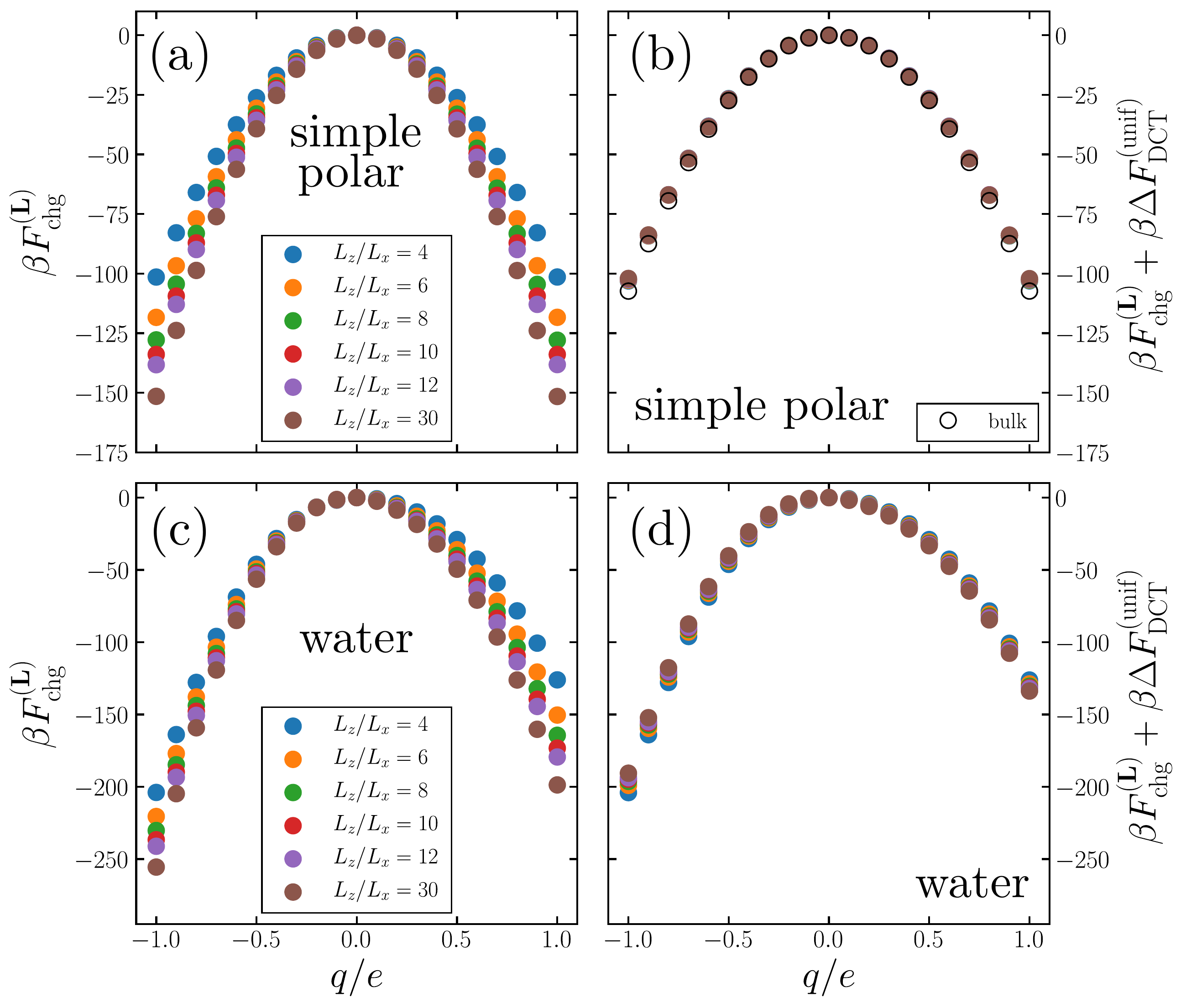}
  \caption{System size dependence of solute charging free energy for
    periodic {\em slab} simulations.  In all cases, the solute's
    position was restrained with a harmonic bias potential whose
    minimum coincides with the slab's center of mass.  The lateral
    dimensions $L_{x}=L_{y}\approx 12.4$\,\AA\, of these simulation
    cells are comparable to the cubic cell length for 64 water or 48
    simple polar molecules (see Fig.~\ref{fig:bulk_results}).  (a)
    Results for the simple polar fluid (with 96 molecules and
    $w\approx 27.5$\,\AA{}) depend strongly on $L_{z}$.  (b) Applying
    the dielectric correction from Eq.~\ref{eqn:DeltaFunif-closedform}
    removes nearly all size dependence for this charge-symmetric
    solvent.  For comparison, we also include the bulk results
    obtained with 48 molecules (empty circles).  (c) Results for SPC/E
    water (with 128 molecules and $w\approx 29.2$\,\AA{}) also depend
    strongly on $L_{z}$. The correction from
    Eq.~\ref{eqn:DeltaFunif-closedform} in this case removes much of
    the observed size dependence, but a systematic dependence on
    $L_{z}$ remains.}
  \label{fig:slab_results}
\end{figure*}

Applying this correction to simulation results for periodic slabs of
the simple polar fluid, we can account almost completely for the
substantial variation of charging free energy $F_{\rm
  chg}^{(\mbf{L})}$ with system size.
Fig.~\ref{fig:slab_results}\,(b) shows results for $L_{x}=L_{y}\approx
12.4$\,\AA{}, slab width $w\approx 27.5$\,\AA, and various aspect
ratios $L_z/L_x$, with the solute located equidistant from the liquid
slab's two planar boundaries.  Corrected free energies for these
systems vary by less than 1.05 $k_{\rm B}T$.

Aqueous simulations exhibit a more complicated size dependence, which
is only partially captured by $\Delta F_{\rm DCT}^{({\rm
    unif})}(\mbf{L};q)$.  Charging free energies corrected according
to Eq.~\ref{eqn:DeltaFunif-closedform} are shown in
Fig.~\ref{fig:slab_results}\,(d) for $L_{x}=L_{y}\approx 12.4$\,\AA{},
slab width $w\approx 29.2$\,\AA, and various aspect ratios $L_z/L_x$,
with the solute placed equidistant from the two interfaces. The
variation of $F_{\text{chg}}^{(\mbf{L})}(q)$ with $L_z$ is greatly
reduced by $\Delta F_{\rm DCT}^{({\rm unif})}(\mbf{L};q)$, but a
systematic size dependence remains, leaving differences as large as
13.4 $k_{\rm B}T$ unexplained.  A breakdown of the uniform surface
charge approximation cannot alone explain this failure, since the
remnant size dependence is different for cations and anions -- a
charge asymmetry that cannot be described by DCT. The source of this
charge-asymmetric finite size effect is considered in the next
section.

For the sake of simplicity, Fig.~\ref{fig:slab_results} presents
results only for the case $d=w/2$, i.e., with the ion situated midway
between the liquid slab's boundaries. In Sec.~\ref{sec:fse-asym} we
show that the corrections derived above are equally accurate for other
values of $d$, provided the solute remains in the liquid slab.

\subsection{Finite size effects due to charge asymmetry}
\label{sec:fse-asym}

The simple polar fluid we have considered is charge symmetric, in the
sense that positive and negative charge is equivalently distributed
within each dumbbell molecule. By contrast, the SPC/E model of water
is charge asymmetric, with positive charges situated farther from the
center of volume exclusion. Such asymmetry can lead to a physically
meaningful discrimination between cations and anions. In simulations
with PBC it can also generate unphysical differences. For example, a
solvent comprising volume-excluding molecules whose internal charge
distribution is spherically symmetric cannot respond to the charging
of a solute; yet a bulk simulation of such particles under PBC can
yield a nonzero charging free energy that is different for
cations and anions\cite{wilson1989comment,harder2008origin}. This
artifact reflects a sensitivity of $\phi_{\rm solv}(\mbf{r}_0)$ to the
existence of interfaces even when they are arbitrarily far away.

Bulk simulations with PBC lack interfaces regardless of periodicity,
so that artifacts in $F_{\rm chg}^{(\mbf{L})}$ persist even in the limit
$\mbf{L}\to\infty$. The slab geometry is an intermediate case, with
explicit interfaces but also artifacts and finite size effects of
charge asymmetry that vanish in the $\mbf{L}\to\infty$ limit.  To
demonstrate this fact, we examine the average electric potential
$\phi_{\rm neut}$ experienced by a {\em neutral} solute in a simulation of
periodic slabs.
The dependence of this `neutral
cavity potential' on system size provides a way to extrapolate
computed charging free energies to the macroscopic limit.

\begin{figure}[tb]
  \includegraphics[width=7.65cm]{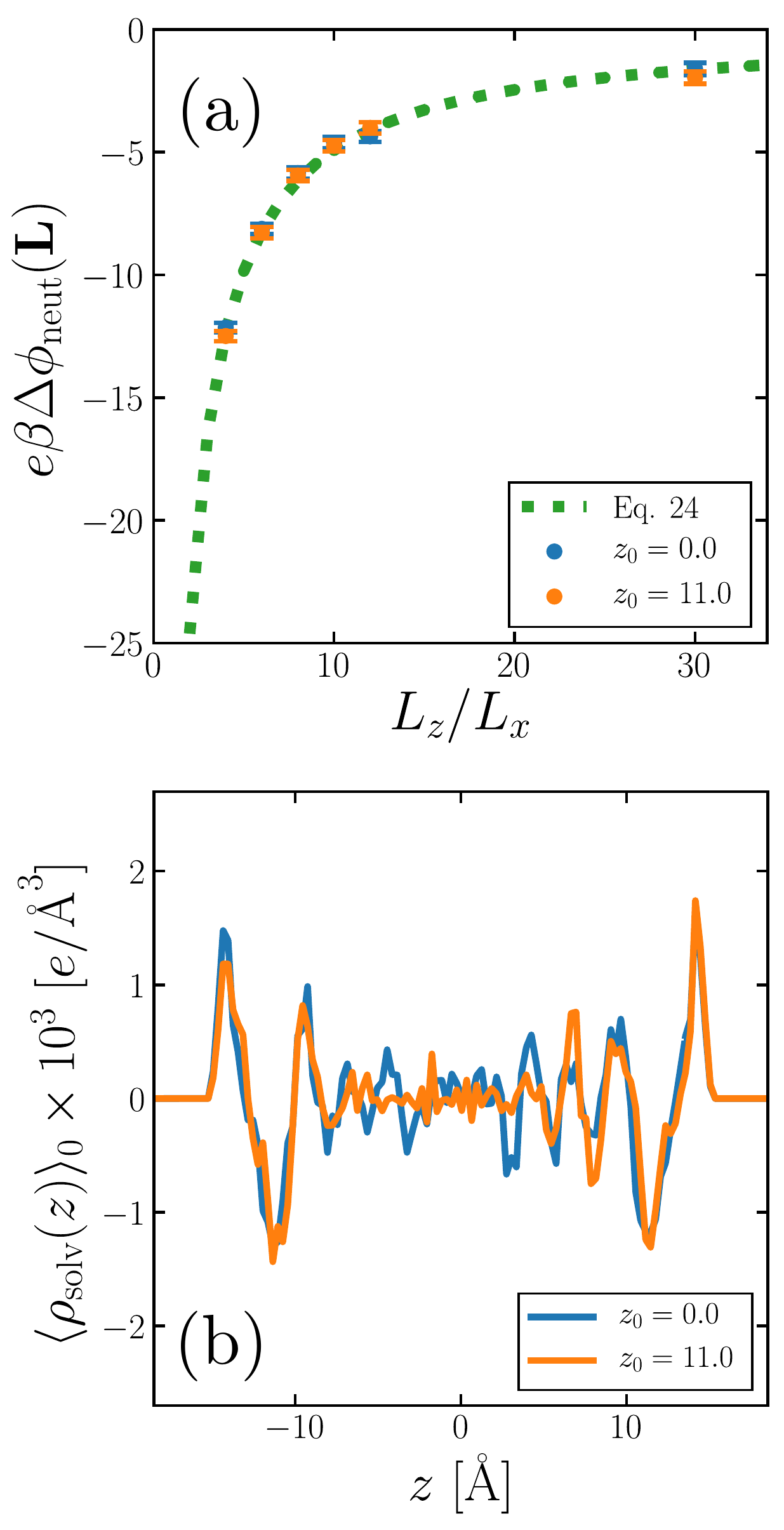}
  \caption{Collective effects of charge asymmetry of an SPC/E water
    molecule.  (a) The neutral cavity potential, relative to that of
    an infinite simulation cell, decays with aspect ratio as predicted
    by Eq.~\ref{eqn:DeltaNeut} (dotted line). Symbols show simulation
    results (with 128 water molecules, $L_{x}=L_{y}\approx
    12.4$\,\AA{}, and $w\approx 29.2$\,\AA) for the solute at the
    center of the slab ($z_{0} = 0.0$) and close to the liquid/vapor
    interface ($z_{0} = 11.0$). Asymptotic values of
    $\phi_{\text{neut}}(\mbf{L};\mbf{r}_{0})$ were obtained by fitting
    $\phi_{\text{neut}}(\mbf{L};\mbf{r}_{0})$ to the form
    $aL_{z}^{-1}+b$, where $a$ and $b$ are free parameters.  (b)
    Average solvent charge density profiles
    $\langle\rho_{\text{solv}}(z)\rangle_{0}$ obtained from
    simulations with a neutral solute. Simulation parameters are the
    same as in (a). Results are shown for two solute positions, in the
    middle of the slab and close to the liquid/vapor interface.}
  \label{fig:phi0_results}
\end{figure}

The neutral cavity potential can be written exactly as:
\begin{equation}
  \phi_{\rm neut}(\mbf{L};\mbf{r}_0) \equiv
  \langle\phi_{\text{solv}}(\mbf{r}_{0})\rangle_{0}^{(\mbf{L})} =
  \sum_{\mbf{b}}\int_{v}\!\mrm{d}\mbf{r}\,
  \frac{\langle\rho_{\text{solv}}(\mbf{r})\rangle_{0}^{(\mbf{L})}}{|\mbf{r}-\mbf{r}_{0}+\mbf{b}|},
\end{equation}
where $v$ denotes a single unit cell of the periodic slab simulation.
$\langle\rho_{\text{solv}}(\mbf{r})\rangle_{0}^{(\mbf{L})}$ is the
solvent's average charge density at position $\mbf{r}$, with a neutral
solute at position $\mbf{r}_0$. Due to volume exclusion,
$\langle\rho_{\text{solv}}(\mbf{r})\rangle$ vanishes inside the solute
(and its periodic replicas). If the solute resides within the slab,
this constraint requires that
$\langle\rho_{\text{solv}}(\mbf{r})\rangle$ depends on $x$ and $y$ as
well as $z$, a significant feature for contributions to $\phi_{\rm
  neut}$ from the primary simulation cell ($\mbf{b}=\mbf{0}$).  Our
interest lies, however, in contributions from more distant unit cells,
which dictate the system size dependence of $\phi_{\rm neut}$. These
distant contributions are much less sensitive to the solvent's
electrostatic inhomogeneity in $x$ and $y$.  We thus expect that
\begin{equation}
  \langle\phi_{\text{solv}}(\mbf{r}_{0})\rangle_{0}^{(\mbf{L})} \approx const + 
  \sum_{\mbf{b}}\int_{v}\!\mrm{d}\mbf{r}\,\frac{\langle\rho_{\text{solv}}(z)\rangle_{0}}{|\mbf{r}-\mbf{r}_{0}+\mbf{b}|}
  \label{eqn:lateral-av}
\end{equation}
is a good approximation, where ${const}$ is independent of $\mbf{L}$,
and the quantity $\langle\rho_{\text{solv}}(z)\rangle_{0}$ has been
spatially averaged over $x$ and $y$.  Fig.~\ref{fig:phi0_results}\,(b)
shows this laterally averaged charge density, obtained from
simulations of SPC/E water with a neutral solute. Results for two
different solute locations (one at the middle of the slab, the other
at the liquid/vapor interface) are nearly indistinguishable. Provided
the solute's cross-sectional area $\pi \sigma^2/4$ is not a
substantial fraction of $L_x L_y$, we expect the laterally averaged
charge density to be insensitive to the solute's presence at all.

Removal of the dependence on the lateral coordinates allows for great
simplification. As shown in Appendix~\ref{sec:generalized-Wigner},
exploiting the symmetries of $\langle\rho_{\text{solv}}(z)\rangle_{0}$
yields
\begin{widetext}
\begin{equation}
  \label{eqn:phi0_far}
  \sum_{\mbf{b}}\int_{v}\!\mrm{d}\mbf{r}\,\frac{\langle\rho_{\text{solv}}(z)\rangle_{0}}{|\mbf{r}-\mbf{r}_{0}+\mbf{b}|} = 
  4\pi\int_{v}\!\mrm{d}z\,\langle\rho_{\text{solv}}(z)\rangle_{0}\bigg[\frac{(z-z_{0})^{2}}{2L_{z}} - \frac{|z-z_{0}|}{2}\bigg]
\end{equation}
\end{widetext}
The {\em difference} quantity
\begin{equation}
  \label{eqn:DeltaNeut_general}
  \Delta\phi_{\text{neut}}(\mbf{L}) = \langle\phi_{\text{solv}}(\mbf{r}_{0})\rangle_{0}^{(\infty)} - \langle\phi_{\text{solv}}(\mbf{r}_{0})\rangle_{0}^{(\mbf{L})}
\end{equation}
can now be written compactly,
\begin{equation}
  \Delta\phi_{\text{neut}}(\mbf{L}) \approx
  -\frac{2\pi}{L_{z}}\int_{v}\!\mrm{d}z\,
  \langle\rho_{\text{solv}}(z)\rangle_{0}z^{2},
  \label{eqn:DeltaNeut}
\end{equation}
where we have eliminated terms that vanish due either to
electroneutrality of the solvent or to the zero net dipole moment of a
pure liquid slab. Numerical results for
$\Delta\phi_{\text{neut}}(\mbf{L})$ are shown in
Fig.~\ref{fig:phi0_results} (a), obtained directly from aqueous
simulations and also from Eq.~\ref{eqn:DeltaNeut}.  They indicate that
neglecting the lateral dependence of
$\langle\rho_{\text{solv}}(\mbf{r})\rangle_{0}^{(\mbf{L})}$ is indeed
a very good approximation. 

\begin{figure}[tb]
  \includegraphics[width=7.65cm]{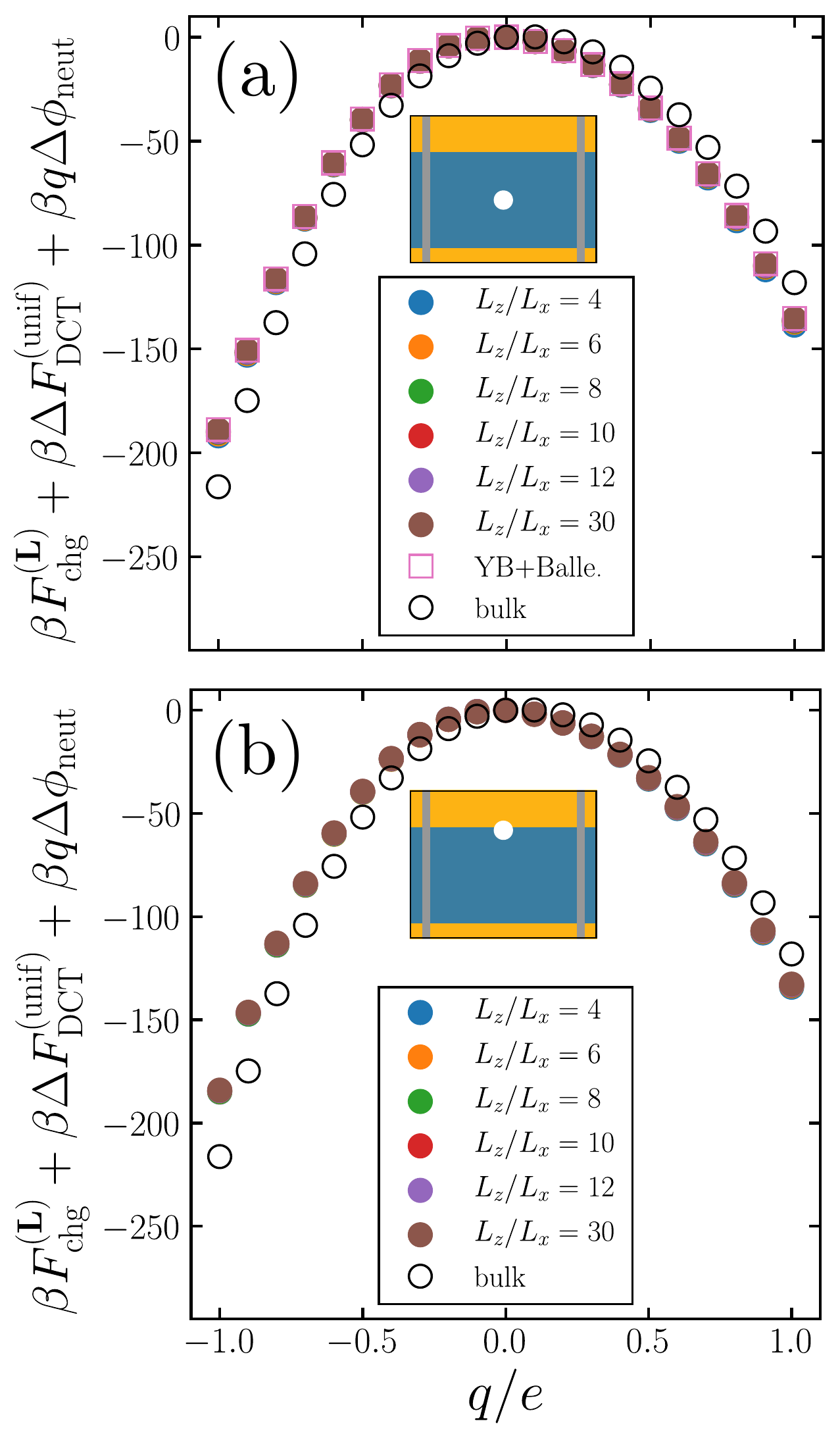}
  \caption{Charging free energies for solutes in periodic water slabs,
    corrected for size dependence of both dielectric response and
    effects of charge asymmetry.  Results are shown for simulations
    with $L_{x}=L_{y}\approx 12.4$\,\AA{} and $w\approx 29.2$\,\AA,
    and for two solute positions: (a) equidistant from either
    interface ($z_{0}=0.0$), and (b) at the air/water interface
    ($z_{0}=11.0$).  For comparison, the empty circles show the bulk
    results obtained with 64 molecules (see
    Fig.~\ref{fig:bulk_results}\,(d)). The empty squares in (a) show
    results from a slab simulation employing the Yeh-Berkowitz method
    \cite{YehBerkowitz1999sjc}, along with the neutralization scheme
    of Ref.~\onlinecite{BalleneggerCerda2009sjc}.}
  \label{fig:slab_fullcorr}
\end{figure}

Eq.~\ref{eqn:DeltaNeut} describes an intrinsic bias on ion solvation
in a system of periodic slabs, one that vanishes in the limit
$L_z\to\infty$. We take this bias, in the absence of any solute charge,
as a baseline for dielectric response, and therefore propose a full
finite size correction for the slab geometry:
\begin{equation}
  \label{eqn:Deltaphisolv_general_surf}
  \Delta F(\mbf{L};q) = \Delta F_{\rm DCT}(\mbf{L};q)
  + q\Delta\phi_{\text{neut}}(\mbf{L})
\end{equation}
Eqs.~\ref{eqn:DeltaFunif-closedform} and~\ref{eqn:DeltaNeut} provide accurate,
inexpensive, and easily implemented approximations for $\Delta F_{\rm
  DCT}(\mbf{L};q)$ and $\Delta\phi_{\text{neut}}(\mbf{L})$,
respectively. Charging free energies corrected according to
Eq.~\ref{eqn:Deltaphisolv_general_surf} are shown in
Fig.~\ref{fig:slab_fullcorr} (a) for an ion at the center of a
periodic slab of liquid water. The remnant size dependence of $F_{\rm
  chg}^{(\mbf{L})}+\Delta F_{\rm DCT}$ evident in
Fig.~\ref{fig:slab_results} (d) is captured almost exactly by the
neutral cavity potential. Eq.~\ref{eqn:Deltaphisolv_general_surf}
accounts equally well for finite size effects observed for an ion
placed at the air/water interface, as shown in
Fig.~\ref{fig:slab_fullcorr} (b).

Several features of the neutral cavity potential are noteworthy.
First, within the approximation leading to Eq.~\ref{eqn:DeltaNeut},
$\Delta\phi_{\text{neut}}(\mbf{L})$ is independent of the solute's
location, even for points well outside the slab. This invariance
suggests that the bias we are correcting is dominated by very distant
contributions, in effect a consequence of omitting each slab's
boundary at $x,y=\infty$ under PBC. Indeed, Eq.~\ref{eqn:DeltaNeut}
resembles the conventional surface potential $\phi_{\rm surf}= 4\pi
\int_{z_{\rm vap}}^{z_{\rm liq}}\!\mrm{d}z\,
\langle\rho_{\text{solv}}(z)\rangle z$, where $z_{\rm liq}$ and
$z_{\rm vap}$ denote locations on either side of a neat liquid/vapor
interface ($z_{\text{vap}}<z_{\text{liq}}$). These potentials are
simply related within a molecular multipole expansion up to quadrupole
order, assuming that the solvent's average polarization field is
nonzero only at the interface and that its quadrupole field is uniform
within the liquid phase. The neutral cavity potential then evaluates
simply to $\Delta\phi_{\text{neut}}(\mbf{L}) \approx \phi_{\rm surf}
w/L_z$.  From this relationship our numerical results for
$\Delta\phi_{\text{neut}}$ suggest a surface potential of roughly
$-0.54$\,V for the SPC/E model, comparable to published values
\cite{sokhan1997free,
  arslanargin2012free,harder2008origin,wick2007effect,lee2007hydration}. Like
the conventional surface potential,
$\Delta\phi_{\text{neut}}(\mbf{L})$ may therefore have little to do
with the molecular physics of solvation, as underscored by the
possibility that $\Delta\phi_{\text{neut}}(\mbf{L})\neq 0$ even for
spherically symmetric solvent molecules. In computing the solvent
potential from a lattice sum method, we effectively adopt the
so-called `P-summation scheme' (see e.g.
Refs.~\onlinecite{reif2016origin}, \onlinecite{darden1998ionic} and
\onlinecite{kastenholz2006computation}) such that the surface
potential contains unphysical contributions from the internal charge
distribution of the solvent molecules.

For bulk liquid simulations such artifacts of PBC are not removed by
extrapolation to the macroscopic limit. For the slab geometry
Eq.~\ref{eqn:lateral-av} indicates that these artifacts are completely
removed as $L_z\to\infty$. Specifically, if the neutral solute resides
in the vapor phase \footnote{We have in mind that the solute is
  sufficiently far from the interface that capillary fluctuations are
  not disrupted.}, $\mbf{r}_0 = \mbf{r}_{\rm vap}$, then
Eq.~\ref{eqn:lateral-av} is exact, with a vanishing constant
term. Integrating Eq.~\ref{eqn:phi0_far} then gives
$\langle\phi_{\text{solv}}(\mbf{r}_{\rm
  vap})\rangle_{0}^{(\mbf{L})}=0$.  In other words, solvent potentials
are properly referenced to the vapor phase when the spacing between
slabs is infinite.  Based on this fact, one convenient way to evaluate
$\Delta\phi_{\text{neut}}(\mbf{L})$ in slab simulations is simply to
compute the average electric potential at a point within the vapor
phase (and far from the interface). Also shown in
Fig.~\ref{fig:slab_fullcorr}\,(a) are the results from a slab
simulation employing the popular Yeh-Berkowitz method
\cite{YehBerkowitz1999sjc} for simulating slab systems, in conjunction
with the neutralization scheme outlined in
Ref.~\onlinecite{BalleneggerCerda2009sjc}. (In this case
$\Delta\phi_{\rm neut} = 0$, and the generalized Wigner potential has
been evaluated numerically in the limit $L_{\rm z}\to\infty$.)  Good
agreement is observed between these results and those obtained with
the \textit{a posteriori} correction derived in this work.

\subsection{Reconciling bulk and slab results}
\label{sec:reconciling}

\begin{figure}[tb]
  \centering
  \includegraphics[width=7.65cm]{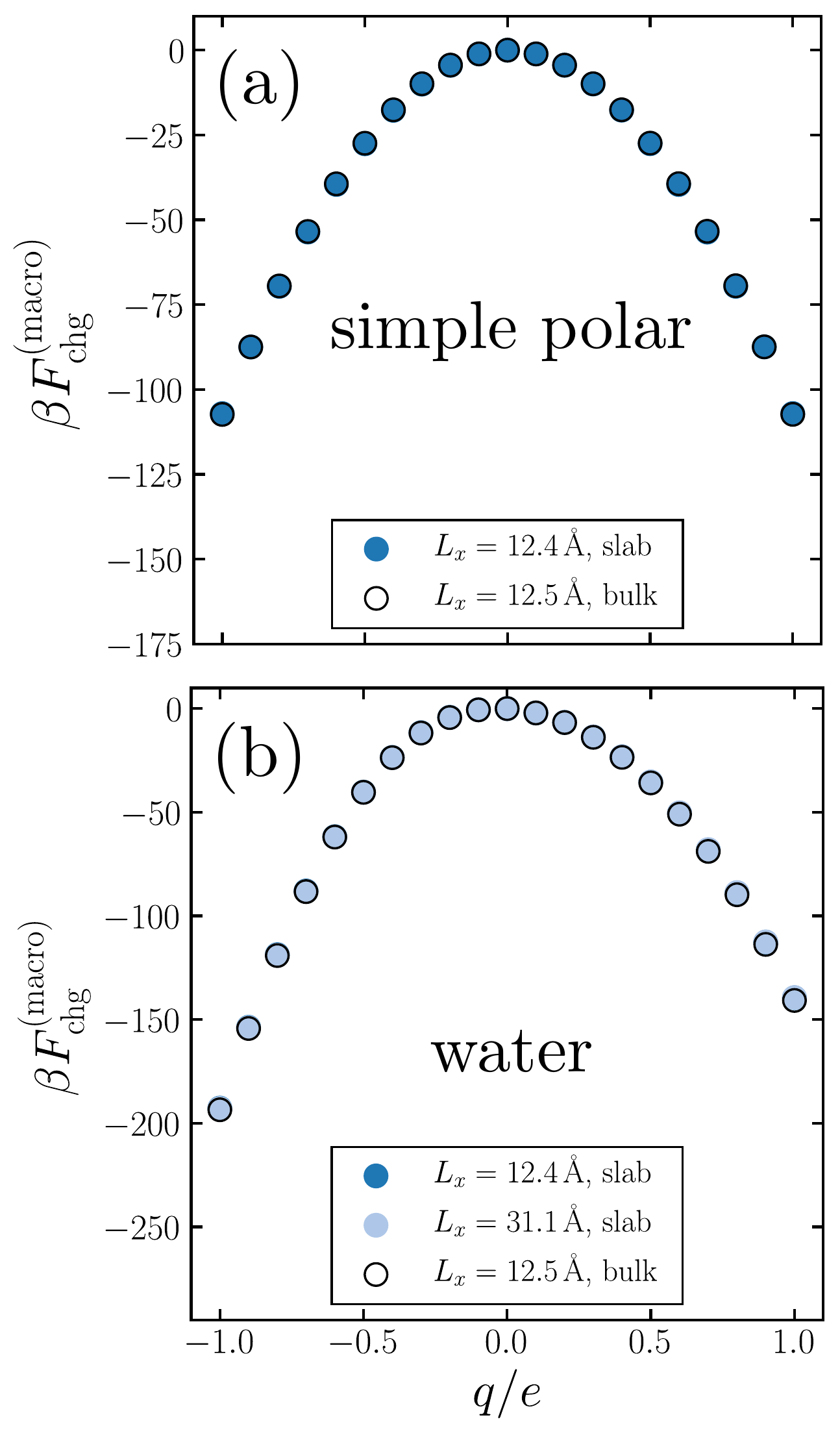}
  \caption{Macroscopic charging free energies, for a solute that is
    distant from any interface, computed from slab and from bulk
    simulations. (a) For the simple polar solvent, slab results (with
    $w\approx 27.5$\,\AA) were corrected according to
    Eq.~\ref{eqn:thickness}, yielding excellent agreement with the
    bulk values.  (b) For water, slab results (with $w\approx
    29.2$\,\AA) were again corrected according to
    Eq.~\ref{eqn:thickness}. Bulk results in this case were also
    corrected, using the surface potential as in
    Eq.~\ref{eqn:Fmacro_bulk}. We also show SPC/E results for a system
    with larger lateral periodicity (see legend), demonstrating
    that effects of finite $L_x$ have been addressed as well as those
    of finite $L_z$. The aspect ratio is $L_{z}/L_{x}
    = 4$ for all slab results shown.}
  \label{fig:bulk_comparison}
\end{figure}

For the periodic slab geometry, with ions at the slabs' centers, the
limit $\mbf{L}\to\infty$ we have considered corresponds to an
individual ion in the middle of an individual slab with infinite
lateral extent.  This system of course still retains a finite
dimension, namely the slab's thickness $w$. Only as $w$ grows should
we expect $F_{\rm chg}^{(\infty)}$ to approach the charging free
energy for a macroscopic solution at infinite dilution. For the simple
polar fluid this remaining finite size effect is evident in
Fig.~\ref{fig:slab_results}\,(b), where deviations are visible between
$F_{\rm chg}^{(\mbf{L})}+\Delta F_{\rm DCT}(\mbf{L})$ for periodic
slab and bulk systems. A correction for finite $w$ can be estimated
from DCT, as described in Appendix~\ref{sec:image_charges}, yielding
an estimate of the macroscopic charging free energy away from any
interface
\begin{equation}
  F_{\rm chg}^{(\rm macro)} \approx F_{\rm chg}^{(\rm slab)}+
  \frac{2q^2}{\epsilon w}
  \ln\bigg(\frac{2}{\epsilon+1} \bigg),
  \label{eqn:thickness}
\end{equation}
where $F_{\rm chg}^{(\rm slab)}$ is the result of periodic slab
simulations with the corresponding finite periodicity correction
$\Delta F_{\rm DCT} + q\Delta\phi_{\rm neut}$ applied.  As shown in
Fig.~\ref{fig:bulk_comparison}\,(a) this adjustment brings slab and
bulk simulation results into good agreement for the simple polar
solvent.

The same finite thickness correction should apply to the aqueous slab
simulations as well.  It is not sufficient, however, to reconcile
differences between slab and bulk results in this case. Indeed, it is
clear from Fig.~\ref{fig:slab_fullcorr} (a) that these deviations are
charge asymmetric, and thus cannot be accounted by DCT.  We argue that
the remaining discrepancy is due not to finite slab thickness, but
instead to artifacts of bulk periodic simulations.  As discussed in
the previous section, such bulk periodic systems lack interfaces even
in the limit $\mbf{L}\to\infty$, which can generate artifacts
correctible only through interfacial considerations. (See
Ref.~\onlinecite{duignan2017electrostatic} for a recent overview of
this issue and of strategies to correct for it.)  Specifically, a
neutral solute at the center of a macroscopic spherical liquid droplet
experiences an average electric potential
\begin{eqnarray}
  \phi_{\rm neut} &=& \int\!\mrm{d}\mbf{r}\,
  \frac{\langle \rho_{\rm solv}(\mbf{r})\rangle_0}{r} \\
  &=& 4\pi \int_0^{r_{\rm vap}}\!\mrm{d}r\, \langle \rho_{\rm solv}(r)\rangle_0 r
  \nonumber \\
  &=&
  4\pi \int_{0}^{r_{\rm liq}}\!\mrm{d}r\, \langle \rho_{\rm solv}(r)\rangle_0 r
  \,+\, \phi_{\rm surf},
  \label{eqn:surf-corr}
\end{eqnarray}
where $r_{\rm liq}$ denotes a point in the liquid far from both the
solute and the liquid/vapor interface.  The first term in
Eq.~\ref{eqn:surf-corr} should be well described by a bulk periodic
simulation as $\mbf{L}\to\infty$, while the latter contribution is
absent. Following the reasoning of the previous section, we account
for the interfaces missing in a bulk periodic simulation by adding $q
\phi_{\rm surf}$ to the computed charging free energy. Our estimate of $F_{\rm chg}^{(\rm macro)}$ from
bulk periodic simulations is thus
\begin{equation}
  \label{eqn:Fmacro_bulk}
  F_{\rm chg}^{(\rm macro)} \approx F_{\rm chg}^{(\rm bulk)}+q \phi_{\rm surf},
\end{equation}
where $F_{\rm chg}^{(\rm bulk)}$ is the result of periodic bulk
simulations with the corresponding finite periodicity correction
$\Delta F_{\rm DCT}$ applied. Numerically integrating $\phi_{\rm
  surf}= 4\pi \int_{z_{\rm vap}}^{z_{\rm liq}}\!\mrm{d}z\,
\langle\rho_{\text{solv}}(z)\rangle z$ for all of our neutral cavity
simulations (using the `P-summation' convention as described above),
we obtain values of $\phi_{\rm surf}$ ranging from $-0.53$\,V to
$-0.58$\,V, comparable to previously reported surface potential values
\cite{sokhan1997free,
  arslanargin2012free,harder2008origin,wick2007effect,lee2007hydration}
for point charge models, as well as our estimate based on
$\Delta\phi_{\text{neut}}$.  Fig.~\ref{fig:bulk_comparison}\,(b) shows
the macroscopic estimates for SPC/E water obtained from
Eq.~\ref{eqn:Fmacro_bulk} with $\phi_{\rm surf} = -0.58$\,V, a value
that yields particularly close alignment of bulk and slab
results. This agreement suggests that the charging free energies in
Fig.~\ref{fig:bulk_comparison}\,(b) are robust assessments of ion
solvation in the physically realistic scenario of a macroscopic
solution with boundaries.  The full macroscopic solvation free energy
would require adding the thermodynamic cost to introduce a neutral
solute, which should not face significant finite size effects.

\section{Discussion}
\label{sec:discuss}

Many simulation studies have examined small ions near aqueous
interfaces using a slab geometry. For system sizes typical of these
studies, we have demonstrated finite size effects that can shift
solvation free energies by 10s of $k_{\rm B}T$. While substantial in
this sense, these effects are unlikely to change many qualitative
conclusions from previous work. In particular, the finite size
corrections of Eqs.~\ref{eqn:DeltaFunif-closedform} and
\ref{eqn:DeltaNeut} do not depend on the ion's location within the
liquid. They are therefore insensitive to a solute's approach to the
interface.  Previously computed density and free energy profiles
should therefore be unaffected for solute positions on the liquid side
of the interface; in this case our corrections simply amount to
resetting the zero of energy.  Our finite size corrections do not
apply well when an ion resides outside the liquid, both because the
uniform surface charge approximation of Eq.~\ref{eqn:DeltaFunif} fails
and because non-dielectric response such as interface deformation is
pronounced. Empirically, we find in this case that aligning simulation
results from different system sizes requires a correction that {\em
  does} depend on the ion's position relative to the interface.
Capturing that dependence is a challenge that likely requires an
approach more nuanced than DCT.

Our results for two model molecular liquids demonstrate that DCT can
predict the system size dependence of interfacial ion solvation
thermodynamics with striking accuracy. This success is remarkable in
light of the length and energy scales involved and the complexities of
interfacial response. The smallest systems we have investigated are
periodic on a scale of just a few molecular diameters, well below the
scale on which the liquid appears continuous. The scale of charging
free energies in Fig.~\ref{fig:bulk_comparison} underscores that the
fluctuations governing solvation of a fully charged ion are
extraordinarily rare (with relative probability $< e^{-100}$) in its
absence. In the case of water, the overall response to charging a
solute is markedly nonlinear. Both in bulk water and in aqueous slabs,
$F_{\text{chg}}(q)$ deviates strongly from the parabolic form
characteristic of linear response. (Most notably, the curvature $d^2
F_{\text{chg}}/dq^2$ differs significantly at positive and negative
$q$, so that anions are far more favorably solvated than cations
\cite{HummerGarcia1996sjc,hirata1988viewing,grossfield2005dependence,bardhan2012affine,lynden1997hydrophobic}.
A contribution linear in $q$, which favors cation solvation, is much
weaker for appreciably charged solutes ($|q| \gtrsim 0.5$)
\cite{reif2016origin}.) The softness of liquid/vapor interfaces and
the altered hydrogen bond network structure at the liquid's boundary
contribute additional nonlinearities that shape ions' association with
the surface
\cite{benjamin1991theoretical,venkateshwaran2014water,OttenSaykally2012sjc,Noah-VanhouckeGeissler2009sjc,kumar2013exploring}. And
yet a continuum linear response theory describes nm-scale
thermodynamic finite size effects to an accuracy of $\lesssim k_{\rm
  B}T$.  From the non-parabolic shape of $F_{\text{chg}}(q)$ we know
that this success of DCT cannot hold on all length scales.  Our
results indicate that charge-asymmetric nonlinear response in such
water models is confined to the very near field of one or two
solvation shells.

Our formulation of finite periodicity corrections for inhomogeneous
geometries is motivated by a specific interest in how liquid water
accommodates solutes near interfaces, and it builds directly on
previous work from the molecular physics community.  But similar
finite size effects have been investigated in other areas, e.g.,
condensed matter physics (see
e.g. Refs.~\onlinecite{makov1995periodic,freysoldt2009fully,komsa2012finite,komsa2013finite,Andreussi2014Electrostatics}). The
results and methods described here may thus be useful in a range of
contexts involving polarization response, such as charged defect
formation in semiconductors, and obtaining reliable reference
potentials in computational electrochemistry
\cite{cheng2014redox,le2017determining}.


\section{Conclusions}
\label{sec:concl}

In this article, we have advanced a physical perspective and a
mathematical framework for assessing the impact of spatial periodicity
on ion solvation free energies computed from molecular simulation.
Our general result Eq.~\ref{eqn:Deltaphisolv_general} is a finite size
correction from dielectric continuum theory, suitable for application
to systems with boundaries where the local polarizability changes
sharply. It requires as input long-wavelength features of induced
surface polarization charge.  For the case of a charged solute in a
dielectric slab under periodic boundary conditions,
taking this boundary charge to be uniform is an excellent
approximation, which yields Eq.~\ref{eqn:DeltaFunif-closedform} as an
easily computable correction.

For solvent molecules that are charge-asymmetric (such as water), an
additional consideration outside the scope of DCT is needed to
describe finite size effects in the slab geometry. Related to the
surface potential, the resulting correction in Eq.~\ref{eqn:DeltaNeut}
removes artifacts associated with the absence of distant interfaces
under PBC.  As $L_z\to\infty$ this absence becomes inconsequential. By
contrast, for periodic bulk simulations the absence of interfaces is
problematic even in the limit $\mbf{L}\to\infty$. Using the surface potential
to effectively restore those interfaces, we find that fully corrected
bulk and slab results approach a consistent macroscopic limit.

We regard these successes as evidence that the electrostatic response
of aqueous environments is well described by DCT down to length scales
little larger than a molecular diameter. Theoretical work by others on
the association of ions with the air/water interface has asserted that
the realism of DCT extends even further in that context
\cite{levin2009polarizable,levin2009ions,levin2014ions,tobias2013simulation,baer2012electrochemical,dos2013surface},
in essence to arbitrarily small scales. The methods we have described
facilitate a careful scrutiny of this hypothesis, which we pursue in
ongoing work.


\begin{acknowledgments}
  We would like to thank Dayton G. Thorpe and Layne B. Frechette for
  many fruitful discussions. The work was supported by the
  U.S. Department of Energy, Office of Basic Energy Sciences, through
  the Chemical Sciences Division (CSD) of Lawrence Berkeley National
  Laboratory (LBNL), under Contract DE-AC02-05CH11231.
\end{acknowledgments}


\appendix

\section{Method of images applied to slab systems}
\label{sec:image_charges}

Calculating the dielectric response to a point charge near a planar
boundary between two semi-infinite polarizable media is a textbook
exercise \cite{JacksonJohnDavid1999Ce/J}. Focusing on solvation near
liquid-vapor interfaces, we take one of the regions to be vacuum
($\epsilon=1$) and the other region $\Omega$ to have a large
dielectric constant $\epsilon$. For a point charge $q$ that is located
in the high-dielectric medium at a point $\mbf{r}_0$, lying a distance
$d$ from the interface, Poisson's equation can be easily solved using
the \emph{method of images}. The average electric field within
$\Omega$ is constructed with an effective external point charge, whose
field is divergence-free inside $\Omega$ and enforces dielectric
boundary conditions at the interface $\partial \Omega$. The electric
potential $\phi(\mbf{r})$ is then specified by the requirement that
$\phi$ vanishes at infinite distance from the real charge. For a point
$\mbf{r}$ within $\Omega$,
\begin{equation}
  \label{eqn:phi_image}
  \phi(\mbf{r}) = \frac{q}{\epsilon |\mbf{r}-\mbf{r}_0|}+
  \phi_{\rm image}(\mbf{r})
\end{equation}
\noindent
where 
\begin{equation}
  \label{eqn:phi_image_semiinf}
  \phi_{\rm image}(\mbf{r}) = 
  \frac{aq}{\epsilon |\mbf{r}-(\mbf{r}_0+2 d \hat{\mbf{z}})|}
\end{equation}
is the potential generated by an image charge $a q$ at a point
$\mbf{r}_0+2 d \hat{\mbf{z}}$ that mirrors the location of the real
charge, and $a = (\epsilon-1)/(\epsilon+1)$. The unit vector
$\hat{\mbf{z}}$ is perpendicular to the interface, pointing outwards from the
dielectric.

\begin{figure}[tb]
  \includegraphics[width=7.65cm]{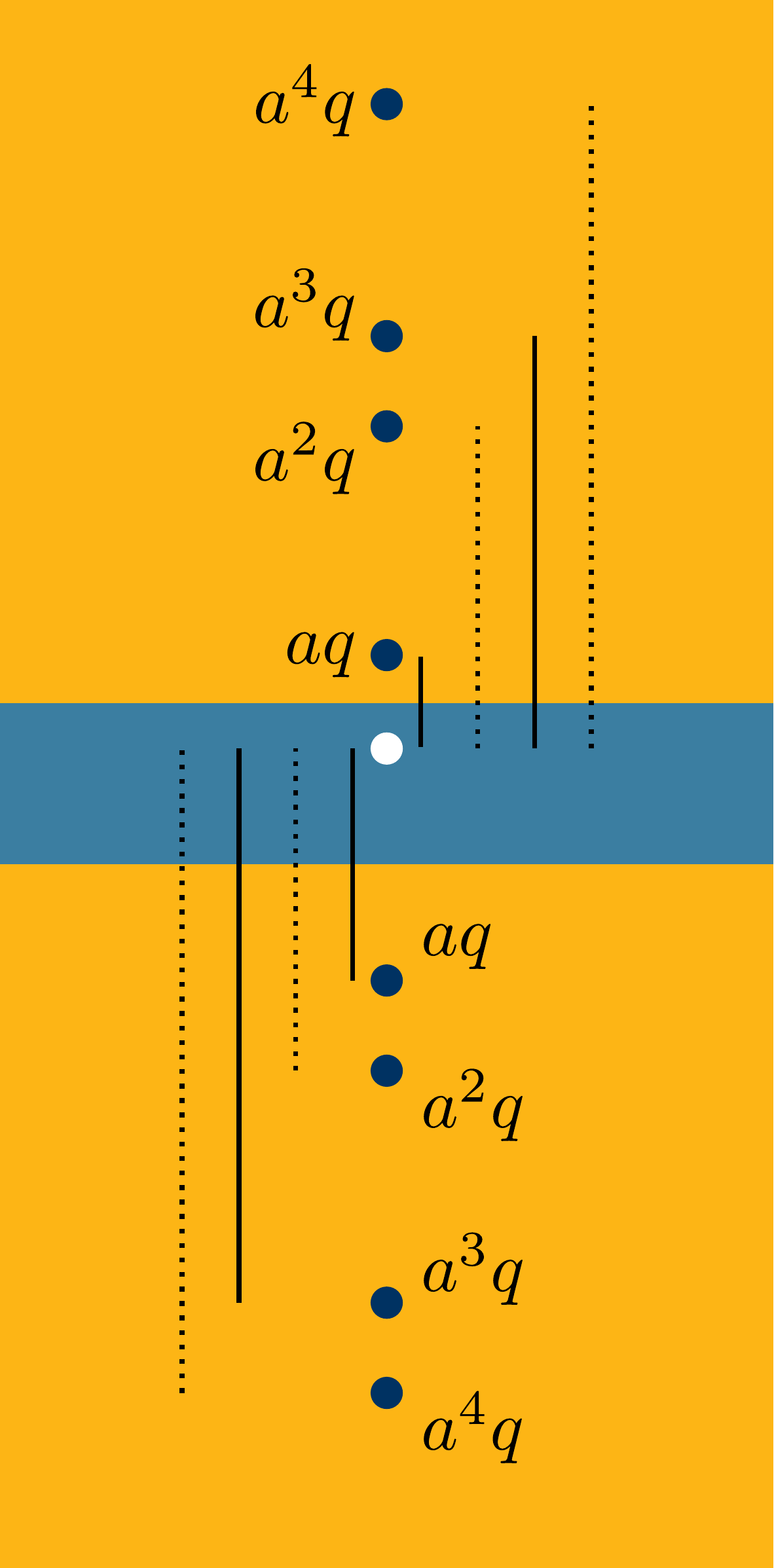}
  \caption{The method of images applied to a point charge in a
    dielectric slab. The slab (light blue) has finite width $w$ and is
    surrounded by vacuum (orange) on either side. The charge $q$
    (white circle) is located a distance $d_{1}$ from the upper
    interface, and a distance $d_{2} = w-d_1$ from the lower
    interface. Dark blue circles represent image charges, which
    collectively generate an electric field in the dielectric that
    satisfies required boundary conditions on both interfaces.  Their
    charges $aq, a^{2}q, a^{3}q,\ldots,a^{n}q,\ldots$ follow a
    geometric progression with increasing distance $\ell_n$ from the
    real charge. For $n$ odd, $\ell_n=(n-1)w + 2d_{1}$ for images
    above the slab and $\ell_n=(n-1)w + 2d_{2}$ below, as shown by the
    solid black lines. For $n$ even, $\ell_n=nw$ on both sides of the
    slab, as shown by the dotted black lines.}
  \label{fig:slab_images}
\end{figure}

\begin{figure}[tb]
  \includegraphics[width=7.65cm]{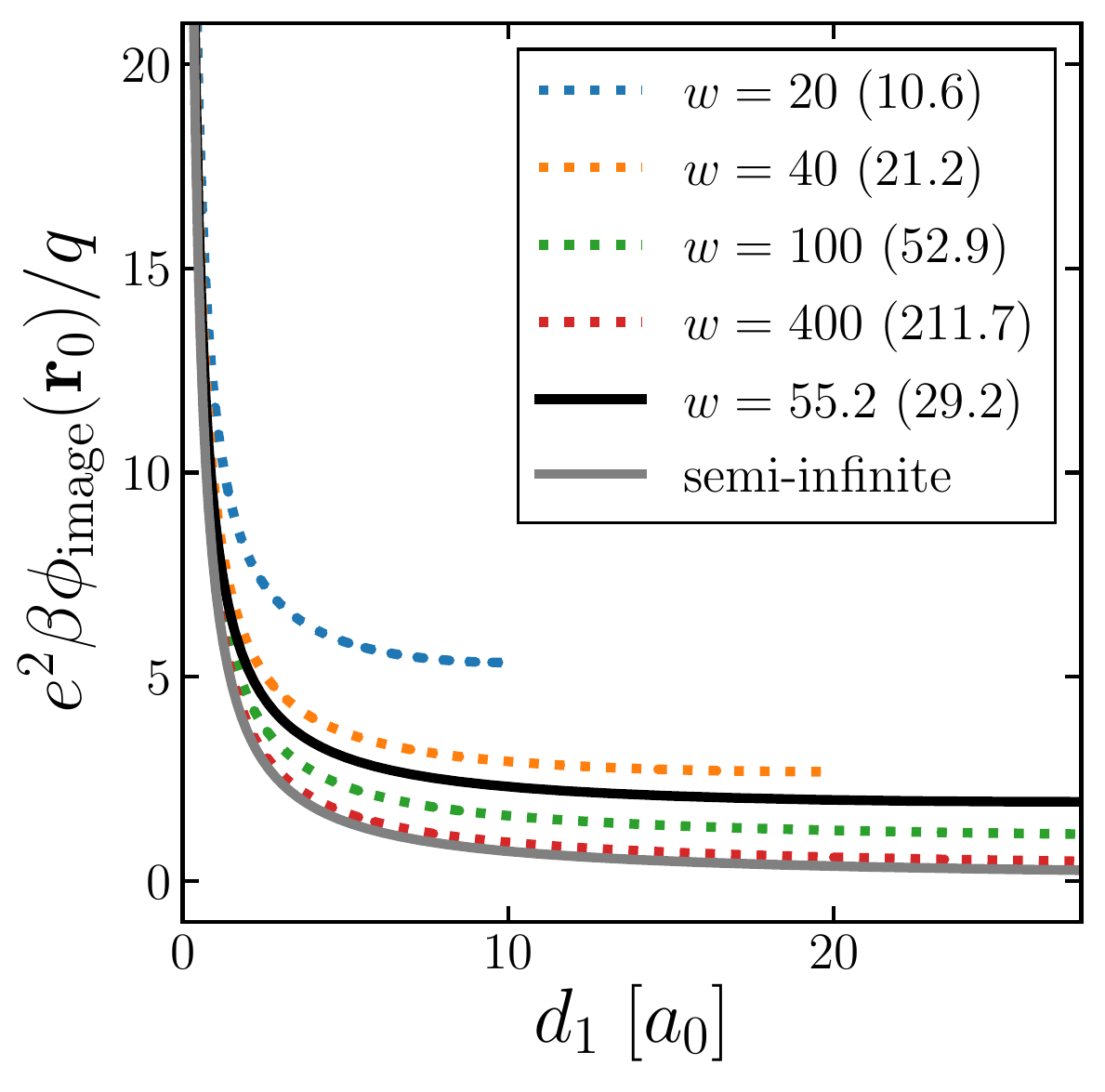}
  \caption{The potential due to image charges at the location of a
    point charge in a slab of width $w$, as given by
    Eq.~\ref{eqn:phi_imag_general}. The point charge is located a
    distance $d_{1}$ from the nearest interface. Both $d_{1}$ and $w$
    are reported in units of Bohr (\bohr); for convenience, $w$ in \AA
    ngstrom is given in parentheses in the legend. The solid black
    line shows the result for the water slab used in the main
    article. The solid gray line shows the result for the
    semi-infinite slab given by Eq.~\ref{eqn:phi_image_semiinf}.}
  \label{fig:phi_images}
\end{figure}

Solvation of a point charge in a dielectric slab, bounded by two
planar interfaces with vapor, can be similarly solved. In this case
satisfying boundary conditions on both interfaces requires an infinite
series of image charges, as depicted in Fig.~\ref{fig:slab_images}.
Here the real charge lies a distance $d_{1}$ from its closest
interface (the top interface in Fig.~\ref{fig:slab_images}), and a
distance $d_{2} = w - d_{1}$ from the other interface (the lower
interface in Fig.~\ref{fig:slab_images}), where $w$ is the slab
thickness. Above the slab reside image charges: (i) $aq, a^{3}q,
a^{5}q,\ldots$ at positions $\mbf{r}_0+2d_{1}\hat{\mbf{z}},
\mbf{r}_0+2(w+d_{1})\hat{\mbf{z}},
\mbf{r}_0+2(2w+d_{1})\hat{\mbf{z}},\ldots$, and (ii) $a^{2}q, a^{4}q,
a^{6}q,\ldots$ at positions $\mbf{r}_0+2w\hat{\mbf{z}},
\mbf{r}_0+4w\hat{\mbf{z}}, \mbf{r}_0+6w\hat{\mbf{z}},\ldots$, where
$\hat{\mbf{z}}$ is an outward normal for the {\em upper} interface.
Below the slab reside image charges: (i) $aq, a^{3}q,
a^{5}q,\ldots$ at positions $\mbf{r}_0-2d_{2}\hat{\mbf{z}},
\mbf{r}_0-2(w+d_{2})\hat{\mbf{z}},
\mbf{r}_0-2(2w+d_{2})\hat{\mbf{z}},\ldots$, and (ii) $a^{2}q, a^{4}q,
a^{6}q,\ldots$ at positions $\mbf{r}_0-2w\hat{\mbf{z}},
\mbf{r}_0-4w\hat{\mbf{z}}, \mbf{r}_0-6w\hat{\mbf{z}},\ldots$.
Since the electric potential must again vanish as $\mbf{r}\to\infty$,
$\mbf{\phi}(\mbf{r})$ is still given by Eq.~\ref{eqn:phi_image},
now with
\begin{widetext}
  \begin{align}
    \phi_{\text{image}}(\mbf{r}) &=
    \sum_{j=0}^{\infty}\frac{a^{2j+1}q}
        {\epsilon |\mbf{r}-[\mbf{r}_0+2(jw+d_{1})\hat{\mbf{z}}]|}
        + \sum_{j=1}^{\infty}\frac{a^{2j}q}
        {\epsilon|\mbf{r}-[\mbf{r}_0+2jw\hat{\mbf{z}}]|}
        \nonumber \\
        & +
\sum_{j=0}^{\infty}\frac{a^{2j+1}q}
        {\epsilon |\mbf{r}-[\mbf{r}_0-2(jw+d_{2})\hat{\mbf{z}}]|}
        + \sum_{j=1}^{\infty}\frac{a^{2j}q}
        {\epsilon|\mbf{r}-[\mbf{r}_0-2jw\hat{\mbf{z}}]|}
        \nonumber \\
    \label{eqn:phi_imag_oneslab}
  \end{align}
\end{widetext}
Finite size corrections in Sec.~\ref{sec:slabs} involve the potential
$\phi(\mbf{r})$ evaluated at the solute's position. For this case of a
single ion in a single slab, $\phi(\mbf{r}_0)$ can be written exactly
in terms of the Hurwitz-Lerch transcendent
$\Phi_{n}^{\text{HL}}(z,b)$:
\begin{widetext}
\begin{equation}
  \label{eqn:phi_imag_general}
  \phi(\mbf{r}_0) = \frac{1}{\epsilon}\phi_{\rm self}
  -\frac{q}{w\epsilon}\ln\big(1-a^{2}\big) +
  \frac{aq}{2w\epsilon}\bigg[\Phi_{1}^{\text{HL}}\left(a^{2},\frac{d_{1}}{w}\right) + \Phi_{1}^{\text{HL}}\left(a^{2},\frac{w-d_{1}}{w}\right)\bigg].
\end{equation}
\end{widetext}
The potential $\phi_{\rm self}$ due to the point ion itself is
singular, but it is independent of the boundaries' geometry and
therefore cancels in all quantities of interest in the main text.  In
the special case that the ion is equidistant from the interfaces
($d_{1} = d_{2} = w/2$):
\begin{equation}
  \label{eqn:phi_image_mid}
  \phi(\mbf{r}_0) = \frac{1}{\epsilon}\phi_{\rm self}
  -\frac{2q}{\epsilon w}\ln\big(1-a\big)
\end{equation}
\noindent In Fig.~\ref{fig:phi_images} we plot the non-singular part
of this potential, $\phi_{\text{image}}(\mbf{r}_{0})$, for various slab
widths.

The case of central interest in this paper -- the periodic slab
simulation geometry depicted in Fig.~\ref{fig:slab_schematic} -- is
more complicated in two respects. First, image charges that determine
the electric field ${\cal E}(\mbf{r}) = -\nabla \phi(\mbf{r})$ do not
follow a simple pattern. Secondly, $\phi(\mbf{r})$ need not vanish as
$\mbf{r}\to\infty$, so that additional information is needed to fully
specify a boundary value problem for the electric potential.  As a
result, the collection of image charges used to solve $\nabla \cdot
{\cal E} = 4\pi \rho({\bf r})/\epsilon$ determine $\phi(\mbf{r})$ only
up to an additive constant. Our approach circumvents this issue (see
Eq.~\ref{eqn:phisolv}) by calculating the potential directly from the
polarization surface charge density $\sigma(\mbf{R})$, which is in
turn completely specified by the electric field:
\begin{equation}
  \sigma(\mbf{R}) = {\bigg(\frac{\epsilon-1}{4\pi}\bigg)}
  \hat{\mbf{n}}\cdot {\cal E}(\mbf{R})
\end{equation}
A full solution in this scheme would first determine image charges
that enforce boundary conditions for ${\cal E}(\mbf{r})$, calculate
the resulting surface charge as a function of position $\mbf{R}$ on
each boundary, and then integrate as per Eq.~\ref{eqn:phisolv} to
obtain the electric potential. This procedure would be tedious but
feasible. For the physical systems we are considering, which feature
an ion inside a high-dielectric solvent that coexists with vapor, an
accurate approximation can be computed much more easily, as explained
below.

\begin{figure}[tb]
  \includegraphics[width=7.65cm]{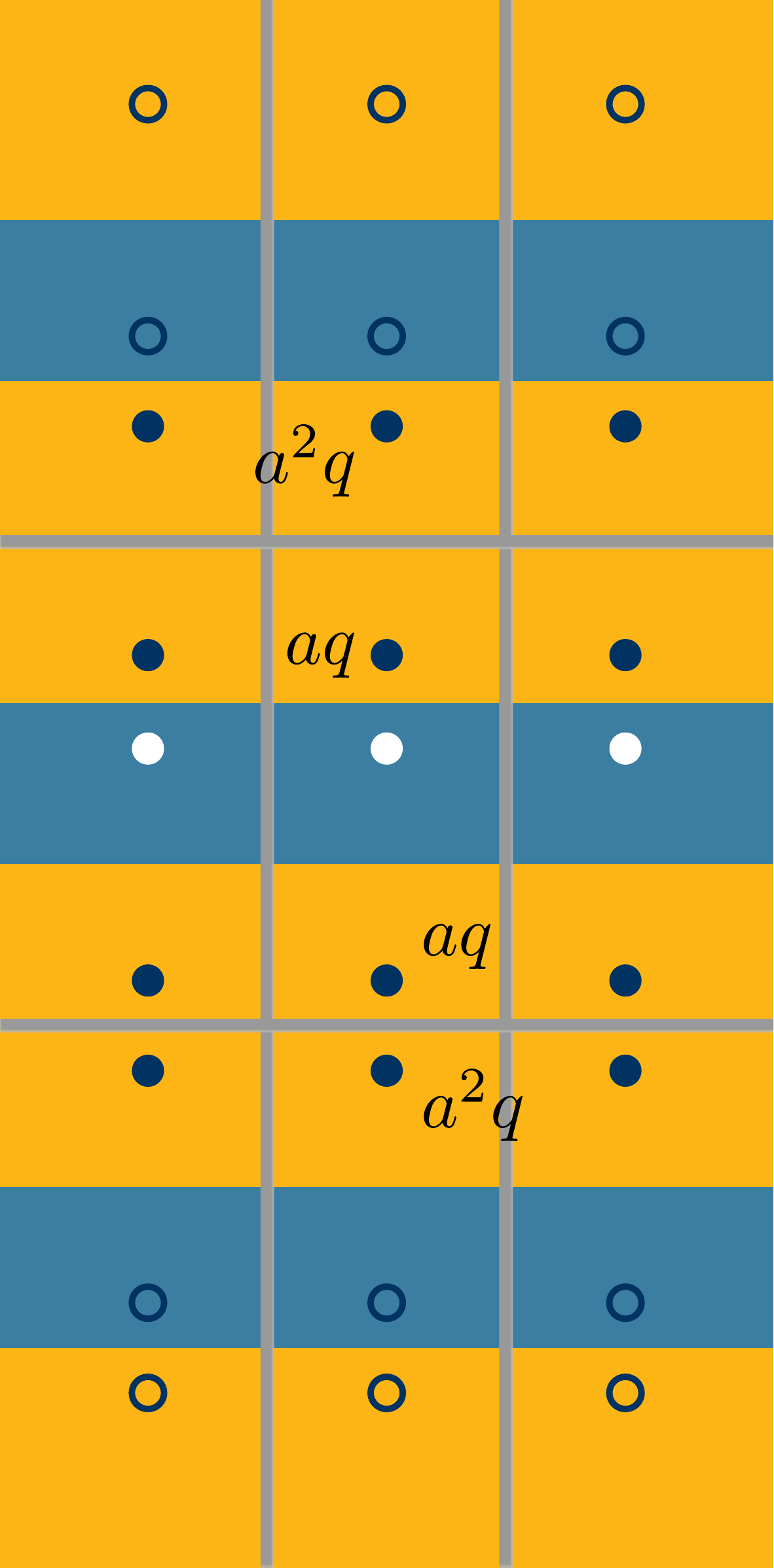}
  \caption{The method of images applied to a periodic collection of
    point charges in a periodic set of dielectric slabs (the periodic
    generalization of Fig.~\protect\ref{fig:slab_images}).  The image
    charges in the vacuum regions immediately above and below the slab
    have the same values and positions as in the non-periodic
    case. Beyond these regions, boundary conditions generate a more
    complicated set of values and locations for the images (depicted
    as open circles). Because $L_z/w$ is an integer in this case, the
    locations are discrete and regularly spaced, and the charge values
    can be written recursively.}
  \label{fig:slab_images_per}
\end{figure}

For a periodic collection of point charges in dielectric slabs, the
method of images gives a series of effective charges exemplified by
Fig.~\ref{fig:slab_images_per}. For a given real charge, one or more
of the nearest images are identical to the non-periodic case of
Fig.~\ref{fig:slab_images} (assuming $L_z \ge 3w/2$).  The charges and
positions of more distant images are complicated by the system's
periodic structure (although simple recursion relations can be derived
when $L_z$ is an integer multiple of $w$.  The infinite collection of
real charges also necessitates introducing a neutralizing background
charge, which generates a series of slabs of image charge density (not
shown). The background charge and its images contribute importantly to
${\cal E}(\mbf{r})$ and thus also to $\sigma(\mbf{R})$. Those
background contributions, however, are spatially uniform in $x$ and
$y$, i.e., they are independent of $\mbf{R}$ on each boundary. We
exploit this fact by splitting $\sigma(\mbf{R})=\bar{\sigma}+\delta
\sigma(\mbf{R})$ into a zero-wavevector component $\bar{\sigma}$ and a
component $\delta \sigma(\mbf{R})$ that varies with $\mbf{R}$ and
integrates to zero.  The average surface charge density $\bar{\sigma}$
can be calculated simply and without reference to image charges, as
described in Sec.~\ref{sec:slabs}. The remainder $\delta
\sigma(\mbf{R})$ receives no contribution from the background charge
or its images.

Results presented in Sec.~\ref{sec:slabs} were obtained neglecting the
lateral variation of polarization surface charge, in effect setting
$\delta \sigma$ to zero. The justifications for this uniform surface
charge approximation are several-fold. First, the most significant
contributions to $F_{\rm chg}$ through $\delta \sigma$ come from the
nearest image charges, which do not depend on $L_z$ and thus cancel in
the calculation of $\Delta F_{\rm DCT}$. Contributions to
$\sigma(\mbf{R})$ from more distant images are smaller in magnitude
and, due to the greater distance, significantly dominated by their
impact on the average charge density, for which $\bar{\sigma}$ already
accounts.  Finally, contributions to the surface polarization charge
from nearby pairs of real and/or image charges that are equidistant
from a boundary are typically of order $1/\epsilon$ for large
$\epsilon$. For example, the physical charge $q$ is largely offset by
an image charge $a q$ opposite the boundary, with the residual $(1-a)q
= 2q/(\epsilon+1) \sim 1/\epsilon$. By contrast, the average charge
density $\bar{\sigma} \sim 1 - 1/\epsilon$ is of order unity.

The uniform surface charge approximation is less defensible when the
solute ion resides instead in the low-dieletric vapor phase. Such a
solute can experience significant solvation forces from multiple
periodic replicas of the liquid slab; contributions from only one of
these replicas will cancel in the calculation of $\Delta F_{\rm DCT}$.
Contributions from nearby charges also do not offset to order
$1/\epsilon$ in this case. For example, the image charge nearest the
interface contributes to $\delta \sigma$ with magnitude
$(\epsilon-1)/(\epsilon+1) \sim 1$, and does not have an offsetting
counterpart opposite the boundary. Empirically, we find that finite
size scaling predicted from this approximation is indeed not closely
followed by simulations. The discrepancy likely has multiple sources.
In addition to $\Delta F_{\rm DCT}^{\rm (unif)}$ poorly approximating
$\Delta F_{\rm DCT}$, we expect that linear dielectric response is a
much cruder model in this case. As small ions leave the liquid phase,
they can deform the interface substantially
\cite{benjamin1991theoretical,benjamin1993mechanism,venkateshwaran2014water},
a response that lies distinctly outside the scope of DCT.

\section{Generalizing the Wigner potential}
\label{sec:generalized-Wigner}

Consider a periodic charge distribution $c(\mbf{r})$ that is (i)
electroneutral, $\int_v d\mbf{r}\, c(\mbf{r})=0$; (ii) non-dipolar,
$\int_v d\mbf{r}\, c(\mbf{r})\, \mbf{r}=0$; and (iii) spatially
uniform in two directions $x$ and $y$, $c(\mbf{r})=c(z)$. The net
potential $\phi(\mbf{r}_0)$ generated by this distribution at
observation point $\mbf{r}_0$ is
\begin{equation}
  \phi(\mbf{r}_0) = \sum_\mbf{b} \int_v d\mbf{r}\,
  \frac{c(z)}{|\mbf{r}-\mbf{r}_0+\mbf{b}|}
\end{equation}
Exploiting the lateral uniformity of $c(\mbf{r})$, this potential
may be written as a Fourier series in just one dimension,
\begin{equation}
  \phi(\mbf{r}_0) = \frac{4\pi}{L_z} \sum_{k_z\neq 0}
  \frac{e^{i k_z(z-z_0)}}{k_z^2} \hat{c}({k_z}),
\end{equation}
where $\hat{c}({k_z}) = \int_{-L/2}^{L/2} dz\, c(z) e^{i k_z z}$,
$k_z=2\pi n/L_z$, and the sum runs over all nonzero integers $n$.  This
sum can be rewritten
\begin{equation}
  \phi(\mbf{r}_0) = \int_{-L/2}^{L/2} dz\, c(z) J(z-z_0),
\end{equation}
in terms of a function $J(z)$ that can be expressed in closed form,
\begin{eqnarray}
J(z) = \frac{4\pi}{L_z} \sum_{k_z\neq 0} 
\frac{e^{i k_z z}}{k_z^2}
= 2\pi \bigg(
\frac{L_z}{6} + \frac{z^2}{L_z} - |z|
\bigg).
\label{eqn:JayZ}
\end{eqnarray}
The second equality in Eq.~\ref{eqn:JayZ} can be shown by twice
differentiating $J(z)$, noting that $L_z^{-1} \sum_{k_z} e^{i k_z z} =
\delta(z)$, and solving the ordinary differential equation that
results.  Doing so is aided by recognizing that $d^2|z|/dz^2 = 2
\delta(z)$, that $J(0)=\pi L_z/3$, and that $J(z)$ is an even function
of $z$.  The assumed symmetries of $c(z)$ allow further
simplification:
\begin{equation}
  \phi(\mbf{r}_0) = 2\pi \int_{-L/2}^{L/2} dz\, c(z)
  \bigg(
\frac{z^2}{L_z} - |z-z_0|
\bigg)
\label{eqn:simple}
\end{equation}

Because the neutralizing charge distributions encountered in
Sec.~\ref{sec:slabs} are uniform in $x$ and $y$, their contributions
to $\Delta F_{\rm DCT}$ can be simplified using Eq.~\ref{eqn:simple}.
In particular, the generalized Wigner potential $\phi_{\rm wig}^*$
that arises for periodic dielectric slabs differs from the
conventional Wigner potential $\phi_{\rm wig}$ only in the form of the
neutralizing background. Defining
\begin{equation}
  \phi_{\rm wig}^* = \phi_{\rm wig} + \Delta\phi^*,
\end{equation}
we can regard $\Delta\phi^*$ as the potential due to periodic,
uniformly charged plates with charge density $\bar{\sigma}$, together
with a compensating background that is uniform {\em outside} the
dielectric slabs and zero inside.
This charge distribution has precisely
the form assumed for $c(\mbf{r})$ above, with
\begin{equation}
c(z) = -\frac{q}{2A}\bigg(1 - \frac{w}{L_{z}}\bigg)\bigg[\delta(z+w/2)+\delta(z-w/2)\bigg]
+ \frac{q}{v}h_{\rm out}(z),
\label{eqn:cz}
\end{equation}
where $h_{\rm out}(z)=1$
outside the dielectric slabs (i.e., for $-L_z/2<z<-w/2$ and for
$w/2<z<L_z/2$) and $h_{\rm out}(z)=0$ inside (i.e., for $-w/2<z<w/2$).
For an observation point inside one of the dielectric slabs ($-w/2 <
z_0 < z/2$), substituting Eq.~\ref{eqn:cz} into Eq.~\ref{eqn:simple}
yields, after simple integration,
\begin{equation}
  \label{eqn:DeltaWigStar}
  \Delta\phi^* = -\frac{\pi q}{3 A L_z^2}(L_z-w)^3
\end{equation}
Note that this result is independent of the observation point $z_0$,
provided it lies within the dielectric.  Note also that $\Delta\phi^*$
vanishes when $L_z=w$, so that the standard Wigner potential is
recovered for a bulk periodic system.
    
\begin{figure}
  \includegraphics[width=7.65cm]{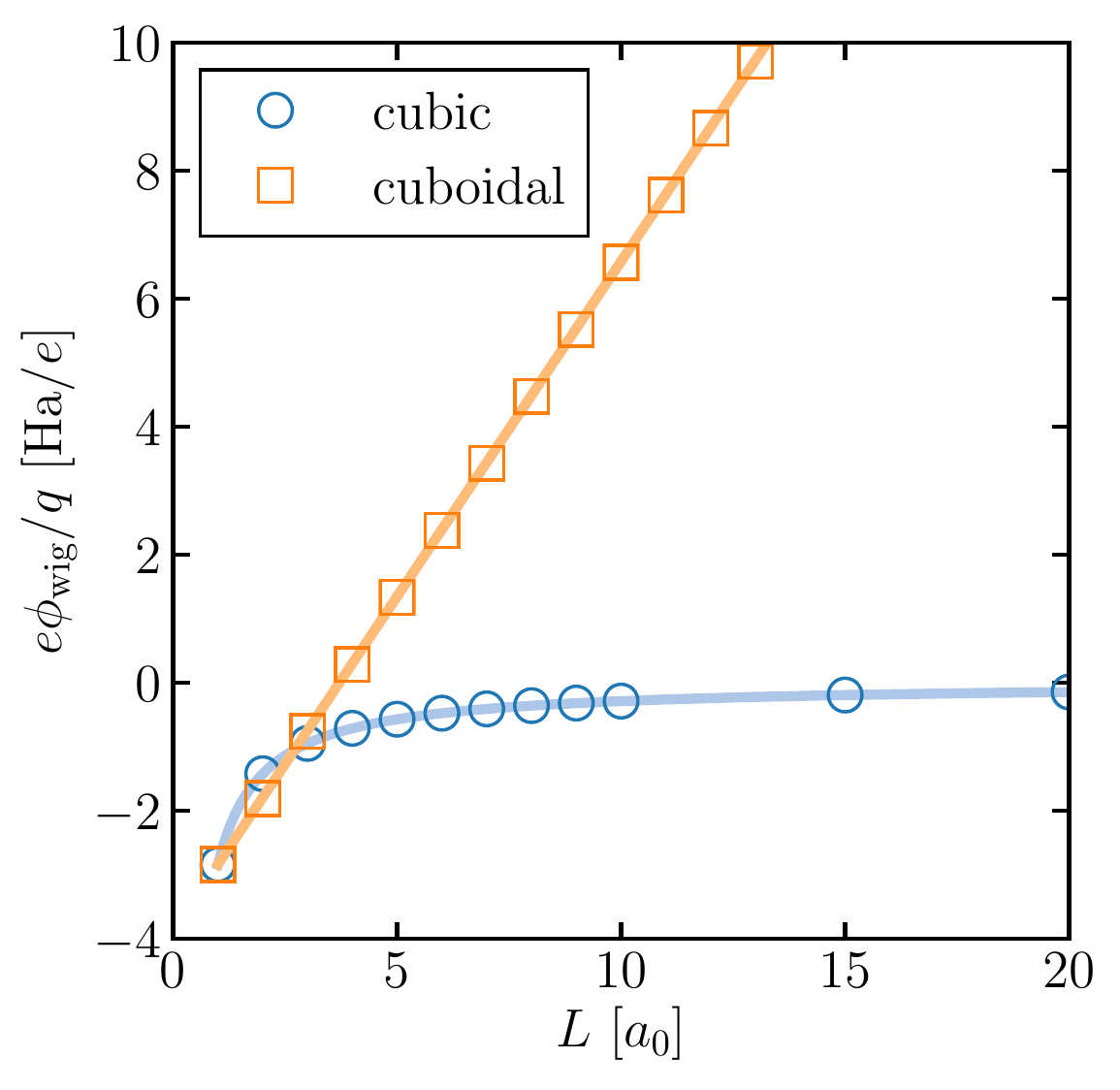}
  \caption{Variation of the Wigner potential, $e\phi_{\text{wig}}/q$
    with system size. Symbols show results of numerical evaluation of
    Eq.~\ref{eqn:numerical-wig}. In the case of cubic cells, $L$ is
    the length of the cell, measured in Bohr (\bohr). For cuboidal
    cells, we set $L_x=1$, and define $L = L_{z}/L_x$ as the aspect
    ratio.  The solid blue line shows the known result
    \cite{NijboerRuijgrok1988sjc} $\phi_{\text{wig}}/q = -2.837297/L$
    for cubic cells. The solid orange line shows the result of linear
    fitting of the cuboidal data, $e\phi_{\text{wig}}/q = 1.0472L -
    3.8845$.}
  \label{fig:phiwig_bulk}
\end{figure}

The conventional Wigner potential, arising from periodic point charges
and a homogeneous background charge $\bar{\rho} = q/v$,
\begin{equation}
  \phi_{\text{wig}} = \sum_{\mbf{b}\neq\mbf{0}} \frac{q}{|\mbf{b}|} -
  \sum_{\mbf{b}}\int_{v}\!\mrm{d}\mbf{r}^{\prime}\frac{\bar{\rho}}{|\mbf{r}^{\prime}+\mbf{b}|}
\end{equation}
is most conveniently evaluated using Ewald summation:
\begin{equation}
  \label{eqn:numerical-wig}
  \phi_{\text{wig}} =
\sum_{\mbf{b}\neq\mbf{0}} \frac{q \,{\rm erfc}(\kappa |\mbf{b}|)}{|\mbf{b}|} +
  \frac{1}{v}\sum_{\mbf{k}\neq\mbf{0}} \frac{4\pi q}{k^{2}}\me^{-k^{2}/4\kappa^{2}}
  - \frac{2q\kappa}{\sqrt{\pi}} - \frac{q\pi}{v\kappa^{2}}
\end{equation}
For cubic simulation cells, the value of $\phi_{\text{wig}}/q \approx
-2.837297/L$ is known to high precision \cite{NijboerRuijgrok1988sjc}.
The finite size corrections described in this paper generally require
$\phi_{\text{wig}}$ for anisotropic cells, which we computed from
Eq.~\ref{eqn:numerical-wig} (with a large value of $\kappa L_x$ that
justifies neglecting the first sum entirely).  For highly anisotropic
cells the magnitude of $\phi_{\text{wig}}$ becomes very large (see
Fig.~\ref{fig:phiwig_bulk}), appearing to diverge linearly with
growing aspect ratio, $e\phi_{\text{wig}}/q = L_x^{-1}(1.0472 L_z/L_x
- 3.8845)$. This divergence precisely matches that of $\Delta\phi^*$
in Eq.~\ref{eqn:DeltaWigStar}, so that the generalized Wigner
potential $\phi_{\rm wig}^*$ remains finite even for very large aspect
ratios.

\clearpage


%

\end{document}